\documentclass[11pt,a4paper]{article}

\usepackage[utf8]{inputenc}
\usepackage[T1]{fontenc}
\usepackage[margin=2.5cm]{geometry}
\usepackage{amssymb}
\usepackage{tipa}
\usepackage{graphicx}
\usepackage{float}
\usepackage{booktabs}
\usepackage{caption}
\usepackage{setspace}
\usepackage[super,sort&compress,numbers]{natbib}
\usepackage[hidelinks]{hyperref}

\newcommand{\eego}{eego\textsuperscript{TM} sports}

\title{\textbf{A 1000-hour EEG--EMG--audio dataset of Japanese speech production}}
\author{
  \parbox{0.9\textwidth}{\centering
  Motoshige Sato$^{1,2,3}$, Ilya Horiguchi$^{1}$, Masakazu Inoue$^{1}$, Kenichi Tomeoka$^{1}$,
  Eri Hatakeyama$^{1}$, Yuya Kita$^{1}$, Atsushi Yamamoto$^{1}$, Ippei Fujisawa$^{1}$,
  Shuntaro Sasai$^{1,4,*}$}\\[0.6em]
  \small $^{1}$Araya Inc., Tokyo, Japan\\
  \small $^{2}$Department of Neurological Surgery, University of California, San Francisco, San Francisco, CA, USA\\
  \small $^{3}$Weill Institute for Neuroscience, University of California, San Francisco, San Francisco, CA, USA\\
  \small $^{4}$Mesotron, Inc., Tsukuba, Japan
}
\date{}

\begin{document}

\maketitle
\vspace{-1em}

\begin{abstract}
\noindent We present a multimodal dataset of 1020 hours of simultaneously recorded scalp electroencephalography (EEG), facial electromyography (EMG), and speech audio from three healthy native Japanese speakers during open-vocabulary overt speech, covert repetition and speech listening. Recordings were acquired in many sessions over several months with three EEG systems of 61--128 channels, an ultra-high-density system (g.Pangolin) and two cap-type systems (g.SCARABEO and \eego{}). Each session provides time-synchronized EEG, facial EMG, and audio, together with speech-event annotations and transcriptions. Although collected with speech decoding as a primary motivation, the dataset also supports work on multimodal signal processing, artifact modeling, longitudinal and cross-device adaptation, and EEG representation learning. Technical validation used power spectral density and event-related potential analyses to characterise spectral profiles and time-locked responses across participants, devices, and tasks. The dataset is available in Brain Imaging Data Structure (BIDS) format from Harvard Dataverse under controlled access for research use. Its speech audio is voice-masked.
\end{abstract}

{\small $^{*}$Corresponding author: Shuntaro Sasai}

\section{Background \& Summary}

Scalp electroencephalography (EEG) remains one of the most accessible non-invasive neural recording modalities and is widely used in neuroscience, clinical neurophysiology, and brain-computer interface research\cite{tudor2005}. Within this landscape, speech is a particularly important target because the loss of voluntary speech severely limits communication, while implanted speech neuroprostheses, although increasingly effective, require neurosurgical intervention\cite{moses2021,willett2023,metzger2023,card2024,anumanchipalli2019,proix2022}. EEG-based speech studies therefore occupy a useful middle ground. Unlike implanted devices, EEG requires no surgery, and unlike functional MRI, it can follow the millisecond time course of speech. These EEG studies motivate communication-oriented applications\cite{dascoli2025,suppes1997,lee2023,kaongoen2022,lopezbernal2022,panachakel2021} and also produce multimodal neural data that can support broader work on signal processing, representation learning, and transfer across devices and sessions.

At the same time, many EEG methods have become increasingly data hungry\cite{kaplan2020,banville2025}. Large public EEG corpora, such as the TUH EEG Corpus, THINGS-EEG, and large-scale M/EEG language datasets, have already accelerated work on clinical modeling, visual representation analysis, and transfer learning\cite{obeid2016,grootswagers2022,defossez2023,schoffelen2019,gwilliams2023}. These resources have also motivated pretrained and foundation-style models for EEG\cite{kostas2021,zhu2023eeg2vec,jiang2024,wang2025,yue2024,elouahidi2025}. However, generalization across participants, sessions, and recording hardware remains a central challenge in EEG research\cite{saha2020,kuruppu2025,heo2025,huang2021,kamrud2021,peterson2021}. Dataset design matters not only in terms of participant count, but also in terms of repeated recordings, device diversity, and the availability of synchronized auxiliary signals.

For speech-related EEG, public data remain relatively limited. Existing datasets have substantially expanded the field, but many are still focused on small prompt sets, imagined rather than overt speech, single-device recordings, or languages other than Japanese\cite{zhao2015,dekker2023,zhang2024,moreira2025,tan2025,zhao2025,chen2026,nieto2022,craik2025}.

Large-scale non-invasive speech decoding has so far relied mainly on speech perception. A study of 175 participants identified perceived speech segments from M/EEG recorded during natural speech listening\cite{defossez2023}. SparrKULee is a repository of EEG from 85 participants listening to natural speech\cite{accou2024}. It supports models that reconstruct the speech envelope or mel spectrogram from EEG, such as ConvConcatNet\cite{xu2024}. These studies show that decoding benefits from large amounts of naturalistic data, but they contain no speech production. Public datasets of overt, covert, or silent speech remain short per participant (Table~\ref{tab:comparison}). This limits within-participant decoding and the study of variability across sessions and devices.

A related question is how far a decoder trained on overt speech can transfer to imagined (covert) speech. Overt speech provides an acoustic reference for aligning neural activity to speech. Recording both modes in the same participant allows this transfer to be tested. Intracranial recordings have shown partly overlapping high-gamma activity during overt and covert word repetition\cite{pei2011} and described the spatiotemporal progression of cortical activity during continuous reading in both modes\cite{brumberg2016}. An electrocorticography model trained on overt productions of eight short sentences could decode covert speech\cite{komeiji2024}. Recordings from motor cortex also revealed a common structure in representations of inner and attempted speech\cite{kunz2025}.

In scalp EEG, the timing of the response can differ between speech modes. In a picture-naming task, Stephan et al. found ERP differences between overt and inner speech during preparation and execution\cite{stephan2020}. Articulatory muscle activity can also contaminate scalp EEG during overt speech\cite{devos2010,porcaro2015}, making simultaneous facial EMG particularly useful in this comparison. In a g.Pangolin study of five words, after adaptive suppression of facial EMG, similar electrodes contributed to overt and covert decoding, but their contributions occurred at different times\cite{sato2026jne}. Among the other datasets in Table~\ref{tab:comparison} that include both overt and imagined speech, only T-MSPD provides both EMG and audio\cite{zhao2025}. Its prompts are limited to ten Chinese characters.

To address the need for longer and more varied recordings of speech production, we assembled JapanEEG (/d\textipa{Z}\ae p\textschwa ni\textlengthmark d\textipa{Z}i\textlengthmark/, from Japanese and EEG). JapanEEG expands our earlier 175-hour dataset\cite{sato2024} to 1020 hours of synchronized EEG, facial EMG, and audio from three healthy participants. The recordings span many sessions and three EEG systems with 61--128 channels. Each participant contributed at least 60 hours of overt-speech recordings with an open vocabulary (Table~\ref{tab:events}). The dataset also includes speech listening for P01 and P02. For P01, the g.Pangolin recordings include both overt speech and covert repetition.

The repeated recordings support within-participant decoding and studies of adaptation across sessions and devices. In one study, a decoder pre-trained on these recordings was adapted to a session seven months later using 30~min of calibration data and was also adapted to a second participant by fine-tuning\cite{sato2026large}. For P01, these g.Pangolin recordings support comparisons between overt and covert speech across many utterances within the same participant and EEG system. The synchronized EEG, EMG and audio also support analyses of articulatory contamination, speech-envelope alignment, and multimodal representation learning. The depth of these recordings complements larger cohorts designed to study generalisation across individuals.

\begin{table}[ht]
\centering
\caption{Comparison of publicly available EEG datasets of speech production. Speech types: O = overt, I = imagined (covert), S = silent. ``Listen'' denotes listening tasks recorded in addition to speech production. Datasets of speech perception only, such as D\'efossez et al.\cite{defossez2023} and SparrKULee\cite{accou2024}, are discussed in the text. \checkmark\ = included; $\times$ = not included. Hours marked with * are estimated from published session counts and durations.}\label{tab:comparison}
\small
\resizebox{\textwidth}{!}{%
\begin{tabular}{llllllllll}
\toprule
Dataset & Lang. & Speech & EEG ch & Participants & Hours & Vocab & EMG & Audio \\
\midrule
KARA ONE\cite{zhao2015} & EN & O/I & 64 & 14 & $\sim$8.2* & 11 prompts & $\times$ & \checkmark \\
DAIS\cite{dekker2023} & NL & O/I & 64 & 20 & $\sim$13.3* & 15 prompts & $\times$ & \checkmark \\
Thinking Out Loud\cite{nieto2022} & ES & I/O & 128+8 & 10 & $>$9 & 4 commands & \checkmark & $\times$ \\
Chisco\cite{zhang2024} & ZH & O/I & 125+6 & 3 & $>$45 & $>$20k sent. & $\times$ & $\times$ \\
Moreira et al.\cite{moreira2025} & EN & Listen/O & 64 & 24 & $\sim$22* & 11+40+40 & $\times$ & $\times$ \\
Tan \& Zhang\cite{tan2025} & ZH & O & 64 & 87 & $\sim$40.4* & 88 targets & $\times$ & $\times$ \\
T-MSPD\cite{zhao2025} & ZH & O/S/I & 64 & 30 & $\sim$24 & 10 chars & \checkmark & \checkmark \\
ChineseEEG-2\cite{chen2026} & ZH & O/Listen & 128 & 12 & 32.4 & Open & $\times$ & \checkmark \\
\textbf{JapanEEG (ours)}\cite{sato2026japaneeg} & JA & O/I/Listen & 61--128 & 3 & $\sim$1020* & Open & \checkmark & \checkmark \\
\bottomrule
\end{tabular}%
}
\end{table}

\section{Methods}

\subsection{Participants}

Three healthy male participants (aged 22--44) took part in this study (Table~\ref{tab:participants}). Throughout this article, participants are labelled P01, P02, and P03, corresponding to the BIDS identifiers \texttt{sub-01}, \texttt{sub-02}, and \texttt{sub-03}, respectively. All were right-handed native Japanese speakers with no history of neurological or psychiatric illness. The study was approved by the Ethics Review Committee of Shiba Palace Clinic (approval number 152506\_rn-35328, 15 June 2023; subsequent approvals 154598\_rn-37020, 21 March 2024; 155344\_rn-37573, 20 June 2024; 156822\_rn-38599, 17 October 2024; and 159354\_rn-40478, 3 July 2025). All experiments were conducted in accordance with national regulations and the World Medical Association's Declaration of Helsinki. Before recording, written informed consent was obtained from every participant for participation in the study and for the recording of EEG, EMG, and audio. Each participant also separately consented to long-term sharing of the recorded data through a data-sharing platform for public benefit, on condition that the data were processed into a form that does not identify individuals. Privacy safeguards in the released dataset include anonymised participant identifiers, voice masking of all released speech audio (see Voice masking), and controlled access under a Data Use Agreement (see Data Availability).

\begin{table}[ht]
\centering
\caption{Demographic characteristics of the study participants. This table summarizes the sex, age, and linguistic background of the individuals ($n=3$) included in the dataset. All participants were healthy Japanese native speakers at the time of data collection.}\label{tab:participants}
\small
\begin{tabular}{llll}
\toprule
Participant & Sex & Age (years) & Description \\
\midrule
P01 & Male & 32 & Native Japanese speaker; healthy adult \\
P02 & Male & 44 & Native Japanese speaker; healthy adult \\
P03 & Male & 22 & Native Japanese speaker; healthy adult \\
\bottomrule
\end{tabular}
\end{table}

\subsection{Experimental paradigm}

\begin{figure}[ht]
\centering
\includegraphics[width=\textwidth]{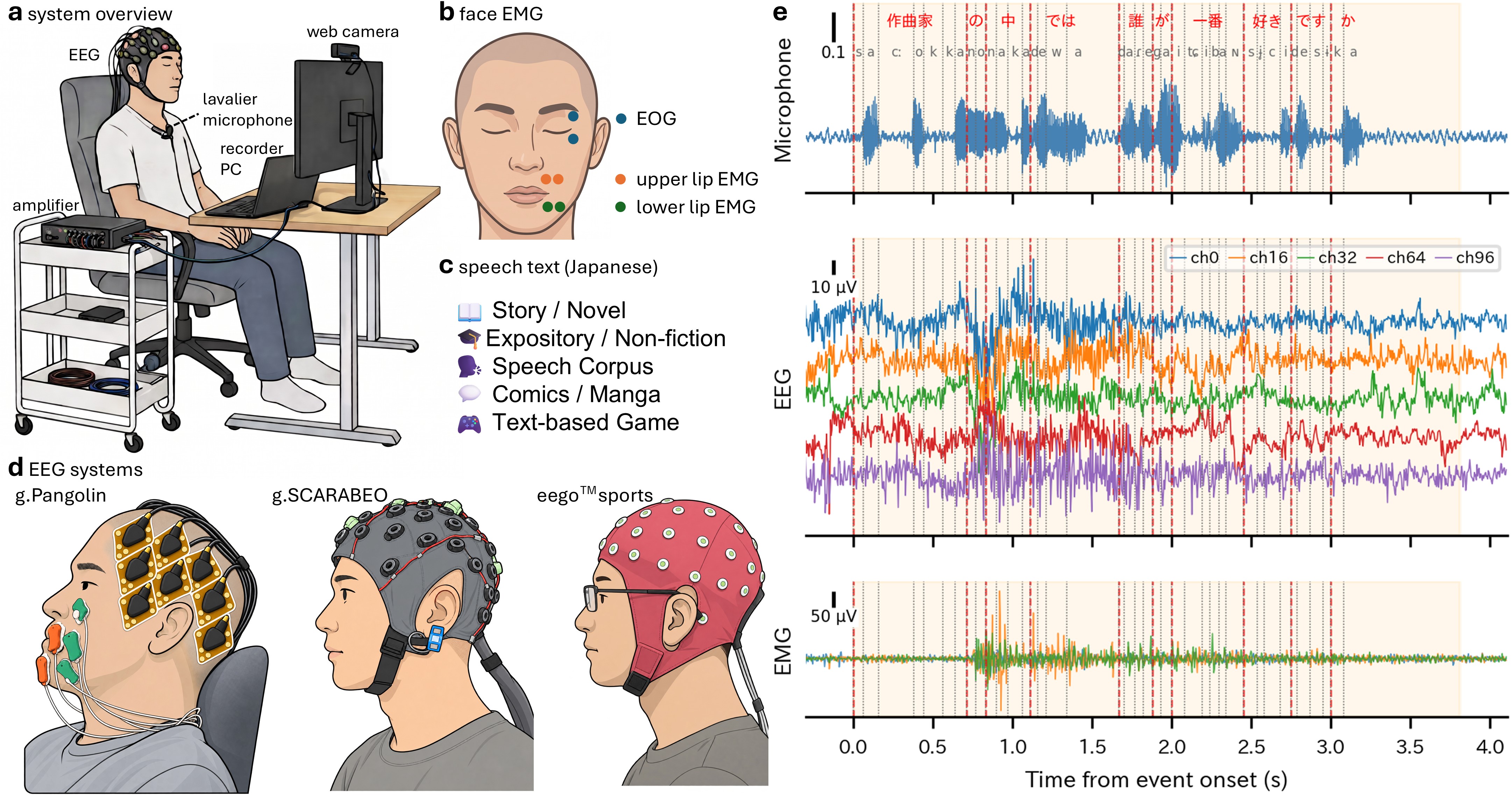}
\caption{Overview of the experimental setup and multimodal speech data recording. (a) Illustration of the experimental environment. The participant is seated in front of a monitor and a web camera, wearing a scalp EEG and a lavalier microphone. The biosignal amplifier also records the microphone signal as the microphone monitor channel, on the same clock as the EEG and EMG. This channel was used to verify and correct the timing of the speech events (see Methods). All signals are recorded via a PC. (b) Face EMG and EOG configuration. Schematic showing the placement of surface electrodes for electrooculography (EOG, blue) to capture eye movements, and facial electromyography (EMG) targeting the upper lip (orange) and lower lip (green) to monitor articulatory kinematics. (c) Speech text categories. The reading task stimuli comprise diverse Japanese linguistic styles sampled from five distinct domains: Story/Novel, Expository/Non-fiction, Speech Corpus, Comics/Manga, and Text-based Game. (d) EEG systems. Illustrations of the three EEG configurations used for dataset collection: g.Pangolin, g.SCARABEO, and \eego{}. (e) Example of synchronized time-series data. Representative data segments aligned to the speech event. The top panel displays the acoustic waveform from the microphone, annotated with the corresponding Japanese text and phone-level transcriptions. Red dashed lines indicate word boundaries, while gray dashed lines indicate phone boundaries. The middle and bottom panels show concurrent EEG signals from representative channels (ch0, ch16, ch32, ch64, and ch96) and facial EMG signals, respectively. Speech intervals, detected using Silero VAD, are indicated by the shaded areas across the panels. The x-axis represents the time in seconds relative to the event onset.}
\label{fig:setup}
\end{figure}

EEG data were accumulated while participants performed overt speech, covert speech and listening tasks (Figure~\ref{fig:setup}). In each session, the participant sat in a chair wearing one of three EEG systems (g.Pangolin, g.SCARABEO, or \eego{}). For overt speech sessions, the participant performed one of the following: (i) playing a text-based television game while reading aloud every line of in-game dialogue, (ii) reading aloud from a book, or (iii) reading aloud sentences from a speech corpus. For covert (imagined) speech sessions, the participant's own voice was first recorded; the participant then listened to this recording and was instructed to reproduce it as imagined speech. We refer to this as the covert repetition task. In continuous listening sessions (P01 and P02), the participant listened through earphones to commercially available audiobooks played on a computer, with no task other than listening. No fixed time limit was imposed on any session; participants continued at their own pace for as long as they were able.

\subsection{EEG and EMG recording}

Three scalp EEG systems were employed across the dataset (Table~\ref{tab:devices}):

\textbf{g.Pangolin} (g.tec; 128 channels). An ultra-high-density system using a 128-channel subset selected from a 1024-channel grid, targeting language and motor cortical areas. Data were sampled at 1200~Hz. The acquisition reference was an electrode on the mastoid, usually the left one. Three channels of facial EMG (upper lip, lower lip, and eye) were recorded simultaneously in a bipolar configuration.

\textbf{g.SCARABEO} (g.tec; 62 channels). A whole-brain EEG system sampling at 1200~Hz. The acquisition reference was an electrode on the mastoid, usually the left one. The earlobe electrodes A1 and A2 were also recorded as reference sites. Three channels of facial EMG (upper lip, lower lip, and eye) were recorded simultaneously in a bipolar configuration.

\textbf{\eego{}} (ANT Neuro; 61 channels). A whole-brain EEG system with Ag/AgCl electrodes arranged according to the international 10--10 system, sampling at 1024~Hz. The final session, recorded on 6 November 2024, was sampled at 2048~Hz. The acquisition reference was CPz. The mastoid electrodes M1 and M2 were also recorded as reference sites. Three channels of facial EMG were recorded simultaneously in a bipolar configuration, with electrodes placed on the eye, upper lip, and lower lip. The lateral side (left or right) was selected per session to accommodate individual skin conditions; within each session all EMG electrodes were placed on the same side.

Before electrode placement, the scalp was prepared and measured to identify the Cz location. For all systems, conductive gel was applied to achieve stable electrode contact. At the start of each recording session with g.SCARABEO and \eego{}, electrode impedances were confirmed to be below 20~k$\Omega$ with the manufacturers' impedance check. For g.Pangolin, signal quality was confirmed by visual inspection of the raw signals before recording.

\begin{table}[ht]
\centering
\caption{Overview of all datasets included in JapanEEG. All participants are healthy adults. \checkmark\ = included. All datasets include simultaneously recorded facial EMG (3 channels, bipolar).}
\label{tab:devices}
\small
\begin{tabular}{lccccc}
\toprule
EEG device & Participants & Hours & Vocab & EMG & Audio \\
\midrule
g.Pangolin (128 ch) & 3 & 731.9 & Open & \checkmark & \checkmark \\
g.SCARABEO (62 ch) & 1 & 134.6 & Open & \checkmark & \checkmark \\
\eego{} (61 ch) & 1 & 153.9 & Open & \checkmark & \checkmark \\
Total & & 1020.4 & & & \\
\bottomrule
\end{tabular}
\end{table}

\subsection{Audio recording}

Speech was recorded with a lavalier microphone clipped to the participant's clothing near the collar. Audio was originally sampled at 48~kHz and downsampled to 16~kHz for analysis. For the open-vocabulary recordings, audio was captured as part of a synchronized video stream (MKV format) using OBS Studio. In addition, the biosignal amplifier recorded two monitor channels on the EEG clock. The microphone monitor channel carries the signal of the same microphone, and the PC audio monitor channel carries the audio output of the computer.

\subsection{Speech event detection}

Speech intervals were identified using the Silero Voice Activity Detection (VAD) model\cite{silero2024}, with processing pipelines tailored to overt speech and listening. Covert speech events were defined from triggers (see below).

For overt speech, adjacent speech segments detected by the VAD model that were separated by a gap of 0.5 seconds or less were merged into a single continuous event. Consequently, short voiceless intervals, such as breath pauses, were included within the total event duration. To mitigate the inclusion of acoustic artifacts and prevent degradation in subsequent transcription accuracy, any detected speech events with a duration of less than 1.0 second were discarded. In continuous listening sessions, we applied the same VAD model and segmentation rules to the stimulus computer's audio output.

The overt speech events were then reviewed by listening to samples of segments with little detected speech. Audio below $-70$~dBFS was not counted as speech, and segments that held only such near-silent audio were removed. The microphone sometimes picked up the voice of the experimenter. Each segment was therefore scored by its speaker similarity to the participant's voice with ECAPA-TDNN\cite{desplanques2020}. The reference embedding was the average of embeddings from clearly voiced segments recorded with the same device. Two of the authors checked by ear all remaining speech segments with a cosine similarity below 0.2 and removed those in which the participant's voice was not heard. The computer's end-of-recording cue is marked as a separate event type. Isolated vocalisations shorter than 1.0 second are marked as short vocalisations. Both can be excluded when the intervals between events are used, for example as periods without speech (see the dataset README).

Overt-speech and listening events were detected in audio files recorded by the computer and placed on the EEG time line. The computer's audio device and the EEG amplifier ran on separate clocks that differed by 15--45~ppm. The resulting drift was about $-1$~ms per minute for g.Pangolin and $-2.5$~ms per minute for g.SCARABEO and \eego{}. A single fixed offset per run would leave the events up to several hundred milliseconds away from the sound by the end of long runs. For every run, we therefore measured the offset between each event and the amplifier's own recording of the same sound. This recording is the microphone monitor channel for overt speech and the PC audio monitor channel for listening (see Audio recording). The offsets were estimated by cross-correlating amplitude envelopes. We used a single broadband envelope without a filter bank, computed as the rectified signal smoothed with a 20-ms moving average and sampled at 1~kHz. The events were then corrected with a robust linear fit of the offsets over each run (Theil--Sen estimator). The median absolute deviation from the fitted line was 0.3~ms for overt speech and 0.8~ms for listening. In 55 runs, single utterances could not be measured reliably, mostly because the microphone monitor signal was weak. In these runs, the offsets were estimated from longer windows or from the onset of facial muscle activity (see the dataset README). The events, word and phone onsets, and voice-masked audio of a run follow the same corrected time map. After clock correction, the lag between the released voice-masked audio and the amplifier microphone had a median of 0~ms across overt-speech fragments (interquartile range, $-3$ to 2~ms). The remaining shift in utterance-event timing on the EEG time line, estimated for each run from its median fragment lag, had a median of 0~ms across runs (interquartile range, 0 to 1~ms, maximum 33~ms).

For the covert repetition task, an event-based timing approach was employed rather than automated VAD. The participant heard a 5-second segment of their own earlier overt speech, which they could replay or skip. The participant was instructed to repeat it mentally during a following 5-second window, started with a key press. The presentation computer sent a trigger at the start of every replay and at the onset and offset of every mental repetition window, and its log recorded which segment was presented. Each replay was defined as a 5-second listening event, shortened when the next trigger came earlier. Each mental repetition window was defined as a 5-second covert speech event, which includes any silent intervals within it. Five covert trials were interrupted or overlapped another trial and are marked separately as covert\_irregular. A trial counted as interrupted when its offset trigger did not come 4.9--5.1~s after the onset. Both event types carry the automatic transcription of the presented segment (Whisper large-v3). Some excerpts replayed from one recording also contained the speech of the text-based game, which had been mixed into the recorded audio. For excerpts that contained both the participant's voice and game audio, we transcribed the participant's speech. Excerpts containing only game audio occurred in 76 \texttt{listening} events and no covert trials. These events are marked \texttt{third\_party\_audio} in the \texttt{reference\_audio\_status} column of the event files (see the dataset README). The event onsets of this task come from the trigger channel, which is recorded on the EEG clock. The clock-drift correction described above was therefore not needed.

The final event-level summary statistics resulting from these segmentation procedures, including event counts and recording durations for each participant--task--device configuration, are summarized in Table~\ref{tab:events}. Additionally, the aggregate event statistics across the entire dataset, grouped by trial type (covert, listening, overt), are presented in Table~\ref{tab:agg}.

\begin{table}[ht]
\centering
\caption{Recording design and event-level summary statistics for each participant--task--device configuration, grouped by participant. ``days'' is the number of recording days and ``recording period'' the first and last month of recording. Together they show which factors (participant, device, task, period) vary together. ``events'' denotes the number of segmented events, and ``event hours'' the cumulative duration of these events. ``rec hours'' indicates the total recording duration corresponding to each configuration prior to event segmentation. Note that the summed event hours (about 540~h) do not equal the 1020~h of recording time. Events cover only the segmented speech or listening intervals, whereas the recording time also includes pauses, other non-speech intervals, and speech segments shorter than 1~s. The covert repetition rows share the same recordings, so their rec hours are counted once in the total. Five interrupted covert trials (covert\_irregular) are not counted. Task names describe the experimental paradigms. The corresponding BIDS task labels are \texttt{speechopen} (overt), \texttt{listeningcovert} (covert repetition) and \texttt{listening} (continuous listening).}
\label{tab:events}
\small
\resizebox{\textwidth}{!}{%
\begin{tabular}{lllrrrrrl}
\toprule
participant & task & device & trial type & events & event hours & rec hours & days & recording period \\
\midrule
P01 & overt & g.Pangolin & overt & 115022 & 160.0 & 346.6 & 104 & 2023-08 -- 2025-02 \\
P01 & overt & g.SCARABEO & overt & 36490 & 75.7 & 134.6 & 41 & 2023-11 -- 2024-02 \\
P01 & covert repetition & g.Pangolin & covert & 20518 & 28.5 & 168.5 & 66 & 2024-03 -- 2024-07 \\
P01 & covert repetition & g.Pangolin & listening & 23148 & 32.1 & 168.5 & 66 & 2024-03 -- 2024-07 \\
P01 & overt & \eego{} & overt & 65264 & 71.5 & 153.9 & 46 & 2024-08 -- 2024-11 \\
P01 & continuous listening & g.Pangolin & listening & 5581 & 6.7 & 10.2 & 6 & 2024-11 \\
P02 & continuous listening & g.Pangolin & listening & 15774 & 18.4 & 23.7 & 7 & 2024-11 -- 2024-12 \\
P02 & overt & g.Pangolin & overt & 49147 & 108.9 & 122.5 & 39 & 2025-05 -- 2025-07 \\
P03 & overt & g.Pangolin & overt & 20289 & 40.3 & 60.3 & 13 & 2024-08 -- 2024-11 \\
\bottomrule
\end{tabular}%
}
\end{table}

\begin{table}[ht]
\centering
\caption{Aggregate event statistics grouped by trial type (covert, listening, overt). ``events'' indicates the total number of events across all configurations, and ``total duration'' the cumulative duration. Event durations are counted up to the end of the EEG recording. The one overt event shorter than 1~s is an utterance that continued after the recording ended.}
\label{tab:agg}
\small
\begin{tabular}{lrrrrrr}
\toprule
trial type & events & \shortstack[r]{total\\duration [h]} & \shortstack[r]{minimum\\duration [s]} & \shortstack[r]{maximum\\duration [s]} & \shortstack[r]{mean\\duration [s]} & SD [s] \\
\midrule
covert & 20518 & 28.5 & 5.00 & 5.00 & 5.00 & 0.00 \\
listening & 44503 & 57.2 & 0.14 & 19.87 & 4.63 & 1.96 \\
overt & 286212 & 456.5 & 0.65 & 20.00 & 5.74 & 4.32 \\
\bottomrule
\end{tabular}
\end{table}

\subsection{Voice masking}

To reduce the identifiability of the participants' voices, all released speech audio was voice-masked. Audio was resampled to 16~kHz and converted with Seed-VC\cite{seedvc2024}, using the offline speech model with 30 diffusion steps and a length factor of 1. Seed-VC extracts the linguistic content of the source speech with a Whisper encoder\cite{radford2023whisper} and regenerates it in the voice of a reference recording. The reference was a fictional voice designed from a text description with Qwen3-TTS\cite{qwen3tts}, reading text written for this project. No real person's voice was used or imitated. Each recording was converted in chunks of at most 20~s, cut at quiet points and joined with 0.5-s cross-fades. The conversion preserved the length of each chunk. We measured each chunk's residual offset by cross-correlating the 1-kHz amplitude envelopes of the original and converted audio and corrected it by shifting the converted chunk. We used the original VAD results and event annotations to define where to keep the converted audio, with up to 0.1~s of padding on either side of a vocalisation (see Speech event detection). We set samples to zero outside these intervals and in any spans marked as the experimenter's voice, silence or the computer's cue.

After correcting the chunk offsets, we compared the original and voice-masked recordings for all 286,212 overt-speech events in the release. For each event, we measured the lag between the amplitude envelopes over the full speech segment. Among the 284,244 segments with a peak correlation of at least 0.5, the median absolute lag was 2~ms and the 95th percentile was 9~ms. To assess variation within a segment, we measured lags in 2-s windows spaced 1~s apart. For the 196,294 segments with at least two windows meeting the same correlation threshold, we calculated the standard deviation of the window lags. The median of these standard deviations was 2.1~ms.

We evaluated re-identification risk with ECAPA-TDNN speaker verification\cite{desplanques2020} in a scenario in which an attacker has separate recordings of the participants and knows the voice-masking method. Enrollment used independently recorded audio of the participants and speech of 50 additional Japanese speakers from the JVS corpus\cite{takamichi2020jvs}. The same voice masking was applied to the enrollment recordings. We report the equal error rate (EER) for speaker verification. The VoicePrivacy Challenge uses this measure to evaluate voice anonymisation\cite{tomashenko2026voiceprivacy}. The EER is the error rate at the similarity threshold where the false-positive rate equals the false-negative rate. In this test, the EER for masked speech was 52\%, close to the chance level of 50\%. For the original speech with unmasked enrollment recordings, it was 14\%. As an additional safeguard, access is controlled under a Data Use Agreement prohibiting attempts to re-identify participants.

We assessed the effect of voice masking on automatic transcription with Whisper large-v3, the primary recogniser used for the released transcriptions. We used 180 overt-speech segments, 60 per participant. To spread the sample across sessions, we took one segment per session in each pass. We included only segments whose original audio could be transcribed. For both audio versions, we compared the automatic transcriptions with manually checked transcripts of the original audio. Because the reference transcriptions record what was heard as kana rather than a standardised kanji spelling, errors were counted on kana readings. Punctuation, symbols and spaces were removed, and the recognised text was converted to its reading with pyopenjtalk. Equivalent kana spellings, such as those of long vowels, were unified in both reference and hypothesis. The character error rate on kana readings was 17.1\% for the original and 25.9\% for the masked speech. This is an increase of 8.8 percentage points, with a one-sided 95\% upper confidence bound of 13.8 points estimated by resampling sessions. Per participant, the rates were 7.5\% and 8.2\% for P01, 4.7\% and 4.8\% for P02, and 26.6\% and 42.6\% for P03. The error rates of P03 were already high for the original speech, because the recogniser partly omitted its fast speech.

\subsection{Power spectral density and Event Related Potential (ERP) analyses}

To assess signal quality, we analysed EEG spectral profiles and event-related responses.

The preprocessed EEG used in these analyses was obtained as follows. The EEG was notch filtered at 50~Hz and its harmonics, band-pass filtered (2--120~Hz), and resampled to 240~Hz. Bad channels were detected in every run on the EEG channels only. We used the criteria of the PREP pipeline\cite{bigdely2015}, namely flat signal, robust amplitude deviation, high-frequency noise, and low correlation with other channels. We added an amplitude criterion. A channel was bad when its standard deviation in the 1--40~Hz band, measured in ten 60-s windows, exceeded five times the median over channels. We excluded bad channels run by run in all analyses. The common average reference was computed over the good channels of each run, because a single broken channel included in the average contaminates all other channels. The mastoid electrodes of \eego{} (M1, M2) and the ear electrodes of g.SCARABEO (A1, A2) are reference electrodes and were not used as EEG channels. Before adaptive filtering, we standardised each channel. For PSD, we used the mean and standard deviation within each run. For ERP, we reduced the influence of outliers during standardisation within each run. We used the median and 1.4826 times the median absolute deviation\cite{robbins2020}. We suppressed facial EMG as in previous work\cite{sato2026jne}. We used a normalized least-mean-squares (NLMS) adaptive filter with a step size of 0.1. For each EEG channel, the filter uses the three EMG channels to predict a component and subtracts it.

For the power spectral density (PSD) analysis, we used Welch's method with 4-s Hann windows and 50\% overlap. For both raw and preprocessed EEG, we computed each channel's PSD within each task segment listed in the event files. We averaged the spectra over the segments of each run. To limit the influence of unusually high-power runs, we took the median PSD across runs for each channel. We did this separately by participant, recording device, and task. For display, we converted the median spectra to dB and averaged them across channels. Each curve was median-centred across the displayed frequency range and vertically offset. We approximated the standard error of the median on the linear PSD scale using the formula for normally distributed values. Applying the delta method gave $(10/\ln 10)\,\sqrt{\pi/2}\times 1.4826\,d/(M\sqrt{N})$ in dB, where $M$ and $d$ are the median and the median absolute deviation of the linear PSD across runs and $N$ is the number of runs of the condition. The shaded band shows the average of these estimates across channels. For the high-frequency comparison of Figure~\ref{fig:psd}, we fitted a line to $10\log_{10}(\mathrm{PSD})$ against $\log_{10}(f)$, where $f$ is frequency in hertz. The fit was made for each channel of the median spectrum over 60--118~Hz, excluding 97--103~Hz around the 100-Hz line-noise peak. We report the median of the channel-wise slopes in dB per decade, where a decade is a tenfold increase in frequency. In interpreting these spectra, we focused on whether the data exhibited the expected broadband decrease in power with increasing frequency and whether power-line noise at 50~Hz and its harmonics was reduced after preprocessing.

For overt speech and listening, we aligned ERPs to the start of voice-activity segments preceded by at least 0.5~s of silence. These onsets match the released speech events, placed on the EEG time line with the drift-corrected time map of each run (see Speech event detection). Voice activity detection marks speech slightly before it becomes audible. On the amplifier's microphone monitor channel, acoustic energy rose about 40~ms after the event onset (median across \mbox{112,748} overt-speech segments of P01 recorded with g.Pangolin). For covert speech, we used the trial onsets in the event files.

For each onset, we extracted an epoch from $-300$ to 800~ms. In overt speech, the ERP departs from the baseline about 150~ms before the event onset (see Technical Validation). This deflection may include articulatory muscle activity. We therefore used the mean from $-300$ to $-200$~ms, before this deflection, for baseline correction. For covert speech and listening, which involve no articulation, we used the mean from $-200$ to 0~ms. For each trial, we measured the maximum absolute amplitude across channels and samples. We rejected trials above three times the median of these values for their condition, removing 6.2--17.3\% of trials per condition. To reduce amplitude differences between trials, we divided each channel's epoch by its own standard deviation. We averaged these epochs within each run and then across runs in which the channel was good. ERP amplitudes are therefore in units of single-epoch standard deviations. We used global field power (GFP) to summarise peak latencies. At each time point, GFP is the standard deviation across channels of the grand-average ERP.

To assess the effect of adaptive filtering, we repeated the ERP analysis for P01 overt speech and listening without the filter. Both analyses used the same good-channel reference, run-wise standardisation, onsets and accepted trials. For each device and condition, we joined the grand-average ERP values across all channels and time points from 0 to 800~ms into one vector per filtering setting. We computed the Pearson correlation between the two vectors and the RMS amplitude of each.

We made Figures~\ref{fig:psd}--\ref{fig:topo} from the original continuous recordings before fragmenting and shuffling them for release (see Data files). We used the \texttt{BAD\_high\_amplitude} marks in the released event files to identify stretches with unusually high amplitudes (see Usage Notes). For recordings released as shuffled files, we mapped these marks back to the time line of the original recordings. We excluded any trials or task segments that overlapped a marked stretch.

To check whether the released files yield similar ERPs, we repeated the ERP analysis on all released files for each condition. We processed continuous runs as in the figures. The shuffled files lack the original run context needed to reproduce run-wise standardisation and adaptive-filter weight history exactly. We processed each shuffled fragment separately and standardised each file robustly, excluding fragments with outlying amplitudes from the calculation. We carried the adaptive-filter weights from one fragment to the next. We used the same onsets, baselines, trial rejection and trial normalisation to reproduce the grand-average ERPs in Figure~\ref{fig:erp}. We computed the Pearson correlation between the reproduced and original ERPs across all channels from $-200$ to 800~ms. The correlations ranged from 0.986 to 0.998 for P01 overt speech with each of the three systems, P02 overt speech, and P01 covert speech. The correlation was 0.91 for P03 overt speech and 0.53 for P01 continuous listening, the condition with the weakest response. These two conditions are released only as shuffled fragments. Some fragments contain less than the required 0.5~s of EEG before the event onset, which reduces the number of usable trials.

\section{Data Records}

The dataset is available from Harvard Dataverse\cite{sato2026japaneeg} under controlled access (see Data Availability) in Brain Imaging Data Structure (BIDS) format\cite{pernet2019}. The BIDS tree is approximately 1.2~TB and is distributed as xz-compressed tar archives (0.37~TB in total).

\subsection{Directory structure}

The tree below shows the BIDS-EEG directory structure after the archives are extracted. Some dated session folders contain only electrode positions (\texttt{electrodes.tsv}) and their coordinate system (\texttt{coordsystem.json}). The shuffled EEG files cover \texttt{speechopen}, \texttt{listeningcovert}, and \texttt{listening}.

\begin{small}
\begin{verbatim}
dataset_root/
  README.md                   (description, access conditions, usage notes)
  code/japaneeg_loader.py     (loader for runs and shuffled segments)
  CHANGES                     (version history)
  dataset_description.json    (dataset metadata)
  participants.tsv/.json      (participant information)
  task-<label>_channels.json  (channel column descriptions; labels:
  task-<label>_events.json     speechopen, listening, listeningcovert)
  task-<label>_acq-<device>_eeg.json
                              (recording parameters per task and EEG system)
  task-speechopen_acq-<device>_recording-vocal_beh.json
                              (description of the voice-masked audio)
  stimuli/
    listening_sources.tsv     (sources of the audio presented for listening)
  sub-01/ ... sub-03/
    ses-YYYYMMDD/             (dated sessions)
      eeg/                    (electrode metadata and continuous files)
      beh/                    (*_recording-vocal_beh.wav for continuous runs)
    ses-shuffled/
      eeg/                    (EDF, channel and event tables, JSON sidecars)
      beh/                    (*_recording-vocal_beh.wav for speechopen
                               *_recording-reference_beh.wav and .json
                               for listeningcovert)
\end{verbatim}
\end{small}

\subsection{Data files}

Raw EEG and EMG signals are provided in European Data Format (EDF). Continuous runs are stored as individual files, and shuffled files combine fragments from multiple runs. The \texttt{\_eeg.json} sidecars describe the recording parameters. The channel files (\texttt{channels.tsv}) give channel names, types, units, descriptions, and indices in the source EDF. Where available, \texttt{electrodes.tsv} gives electrode positions and \texttt{coordsystem.json} defines their coordinate system. The participant table (\texttt{participants.tsv}) gives age, sex, and handedness.

Readings of copyrighted texts, the covert repetition task and audiobook listening are released as shuffled fragments. Each run is divided into fragments around speech or task events and additional fragments covering the remaining recording. For pauses longer than 2.5~s between events, the EEG margins on either side are chosen independently between 0.5 and 1~s. Pauses of 2.5~s or less are divided between the adjacent event fragments at a random point, so these margins can be shorter than 0.5~s. All fragments of a participant, task and EEG system are concatenated in random order. In shuffled files, the mean of each fragment is removed separately from each signal channel. Trigger and status channels retain their original values.

Overt speech is distributed as voice-masked audio (16~kHz mono WAV, see Voice masking). The original speech waveforms are withheld. In the covert repetition task, the 16~kHz reference WAV contains voice-masked excerpts where available. The audio in a listening interval represents the replayed excerpt, and the audio in a covert interval is a reference to the speech the participant was asked to imagine. The first microphone monitor and the PC audio monitor in the EDF carry the amplitude envelopes of the original microphone signal and the played audio, respectively, low-pass filtered at 30~Hz. A second microphone monitor channel, where present, carries the voice-masked waveform aligned to the EEG time line or is set to zero when that waveform is unavailable. For commercial audiobooks, the stimulus waveform is withheld. The file \texttt{stimuli/listening\_sources.tsv} lists the works, editions, narrators and store pages.

Event files (\texttt{events.tsv}) record onsets, event types and transcriptions, including automatic transcriptions of the presented audiobook speech. Isolated vocalisations shorter than 1~s and computer end-of-recording cues are marked as \texttt{short\_speech} and \texttt{cue} events, respectively. The \texttt{value\_asr} column contains automatic transcriptions from Whisper large-v3 via faster-whisper, with kotoba-whisper-v2.0-faster used for known failure patterns such as stock subtitle phrases or unusually short outputs. The transcriptions were generally not manually verified. For readings of public-domain texts from Aozora Bunko, \texttt{value} contains the source passage matched by its reading, and \texttt{transcript\_source} identifies that passage. Otherwise \texttt{value} equals \texttt{value\_asr}. Montreal Forced Aligner provides separate intervals for words and phones (\texttt{word} and \texttt{phone} for automatic transcriptions, and \texttt{word\_ref} and \texttt{phone\_ref} for matched source text). Alignment uses language-specific acoustic models and dictionaries, principally \texttt{japanese\_mfa}, with phones given in the corresponding IPA-based phone set. In audiobook listening, the intervals follow the presented speech. In the covert repetition task, they give timing within the reference excerpt for both the listening and covert intervals. The \texttt{alignment\_flag} column identifies unreliable intervals. A phone longer than 1~s triggers \texttt{long\_phone} for all words and phones in its utterance. This typically indicates speech missing from the transcription. If the final phone extends into a following breath or noise, the final word is flagged \texttt{final\_phone\_long}.

\section{Technical Validation}

Technical validation focused on basic checks of signal quality.

\begin{figure}[ht]
\centering
\includegraphics[width=\textwidth]{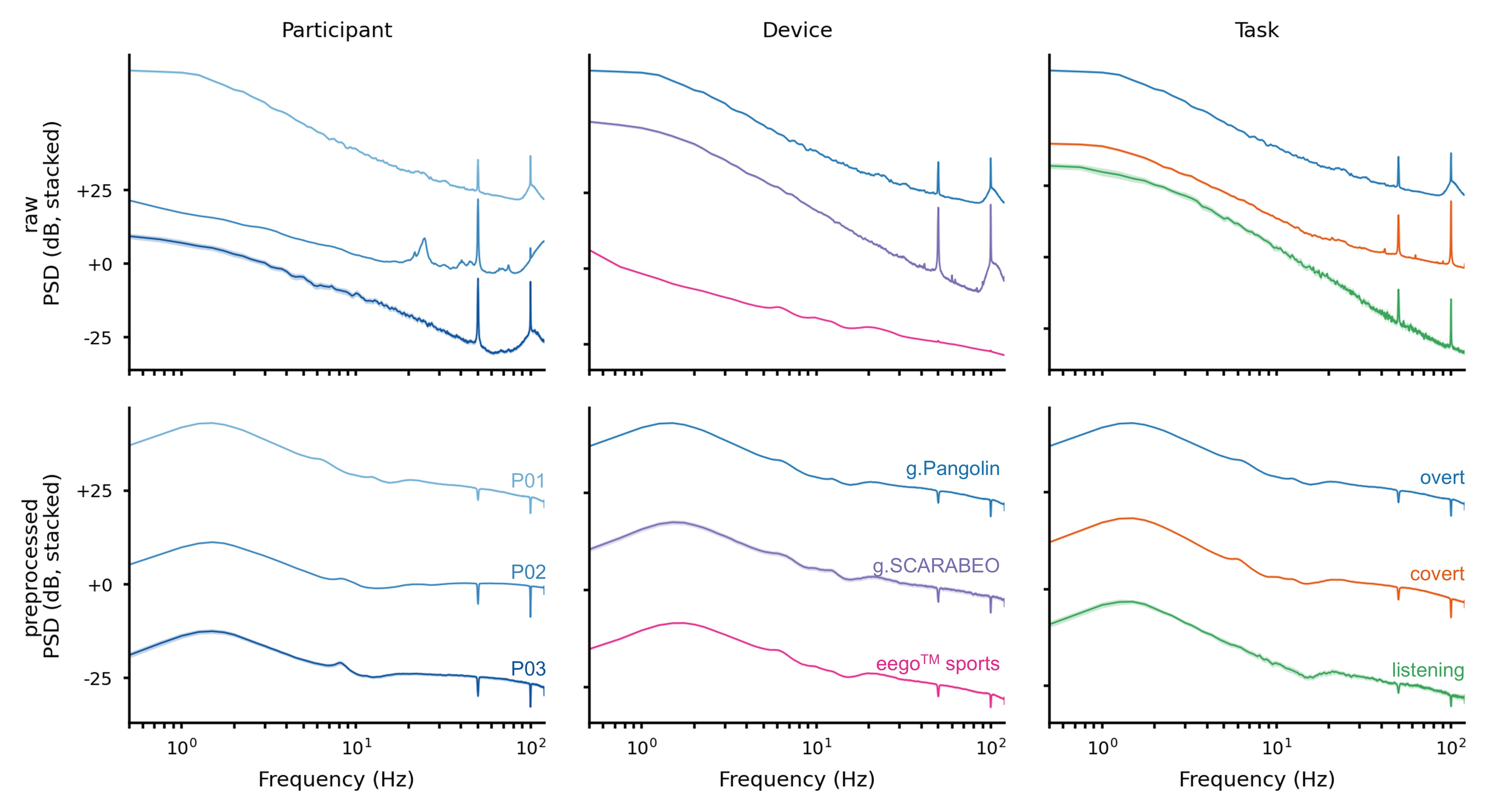}
\caption{Power Spectral Density (PSD) of raw and preprocessed EEG data during speech production and listening. Solid lines show the channel average of the median PSD over runs in dB. Shaded bands show the channel average of the approximate standard error of the median in dB (see Methods). Curves are median-centred and vertically offset. The top row displays raw EEG data. The bottom row shows preprocessed EEG data, after notch filtering, band-pass filtering at 2--120~Hz, common average reference over the good channels of each run, and adaptive filtering, with bad channels excluded. Columns represent comparisons across Participant (left), Device (middle), and Task (right). To ensure comparability, specific conditions are held constant for each column: the Participant column displays data for the overt task using the g.Pangolin device; the Device column displays data for participant P01 during the overt task; and the Task column displays data for participant P01 using the g.Pangolin device.}
\label{fig:psd}
\end{figure}

\textbf{Power spectral density.} To assess the overall signal quality of the recorded EEG, we computed the power spectral density (PSD) across participants, recording devices, and task conditions (Figure~\ref{fig:psd}). The spectra generally decreased with frequency, with an upward trend at high frequencies in the raw overt-speech spectra described below. The preprocessing pipeline attenuated non-neural artifacts, including line noise visible as sharp peaks in the raw spectra, while preserving the broadband spectral structure. Small local peaks near 8~Hz were visible in some preprocessed spectra, most clearly for P03 during overt speech. Between 60 and 118~Hz, the raw spectra during overt speech with g.Pangolin rose with frequency in all three participants. The slope of the median spectrum was $+5.7$, $+34.5$ and $+26.6$~dB per decade for P01, P02 and P03. In P01, the raw spectra fell with frequency during covert speech and listening ($-11.2$ and $-28.0$~dB per decade). We then compared speech with the pauses between utterances within the same runs. For each run, all speech intervals were pooled into one spectrum. All pauses of at least 1.5~s were pooled into another, after removing short vocalisations, coughs and cue sounds. We used all runs whose pauses gave at least 20 analysis windows of 0.5~s (368, 437 and 48 runs of P01, P02 and P03). The median spectrum rose during speech and fell during pauses in all three participants (P01 $+4.7$ vs.\ $-17.7$, P02 $+35.5$ vs.\ $-11.9$, P03 $+22.1$ vs.\ $-21.0$~dB per decade). The slope was higher during speech than during pauses in 364 of 368, 429 of 437 and 48 of 48 runs (Wilcoxon signed-rank test, $p<10^{-14}$ in each participant). Speech and pauses come from the same runs, so day, device and electrode fit are identical. The rise therefore accompanies articulation. It was common to all channels. This is consistent with muscle activity entering through the shared reference, rather than with focal electrode or impedance problems. Muscle activity has a spectrum that extends to high frequencies\cite{goncharova2003,whitham2007}. After common average referencing, the median spectrum during speech no longer rose with frequency. Its slope changed from $+4.7$ to $-7.2$ in P01, from $+35.5$ to $-2.8$ in P02 and from $+22.1$ to $-9.2$ in P03 (dB per decade). The coherence between the EEG and the facial EMG over the same range agrees with this. With the reference electrode used during recording, the magnitude-squared coherence was 0.28--0.62. After common average referencing, it fell to 0.005--0.016, close to the chance level of 0.002--0.012. We estimated the chance level by pairing each EEG window with the EMG of another window. After preprocessing, all spectra fell with frequency between 60 and 118~Hz. The slope of P02 ($-3.9$~dB per decade) remained shallower than those of P01 and P03 ($-9.5$ and $-8.1$). A higher high-frequency noise floor in P02 could contribute to this difference, but these comparisons do not establish its cause. High-frequency activity in these recordings, particularly during overt speech, should not be interpreted as neural gamma activity without further artifact assessment.

\begin{figure}[ht]
\centering
\includegraphics[width=\textwidth]{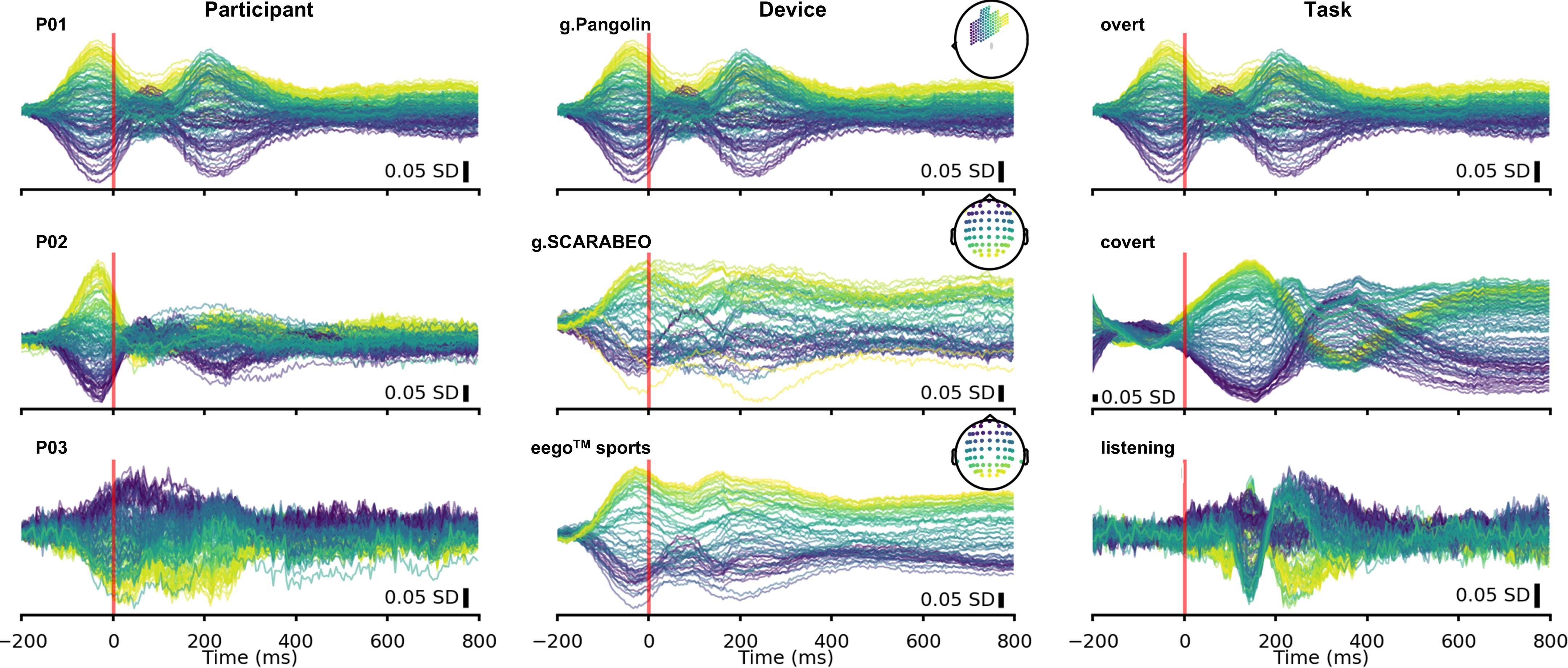}
\caption{Event-Related Potentials (ERPs) validation. ERPs are shown for preprocessed data (Methods). Preprocessing consisted of notch filtering, band-pass filtering at 2--120~Hz, common average reference over the good channels of each run, and adaptive filtering for EMG suppression. Bad channels were excluded, and trials with extreme amplitudes were rejected. Amplitudes are in units of the standard deviation of single epochs (Methods). Columns display the average waveforms grouped by Participant (left), Device (middle), and Task (right). To allow for direct comparison, specific parameters are held constant in each column: the Participant column shows data for the overt task using the g.Pangolin device; the Device column shows data for participant P01 during the overt task; and the Task column shows data for participant P01 using the g.Pangolin device. Individual colored lines correspond to distinct EEG channels. Their colors and positions follow the electrode layouts in the middle column, shown in a left lateral view for g.Pangolin and in a top view for g.SCARABEO and \eego{}. The solid red vertical line at 0~ms marks the event onset. For overt speech and listening, this is the onset of the speech event in the released event files, and the acoustic energy rises about 40~ms later. For covert speech, it is the trial onset. Baselines were $-300$ to $-200$~ms for overt speech, before the pre-onset deflection, and $-200$ to 0~ms for covert speech and listening (Methods). Black vertical scale bars indicate 0.05~SD (standard deviation of single epochs). The vertical axis ranges vary between panels. Bars are placed in the bottom-right corner of each panel, except for covert speech, where the bar is in the bottom-left corner to avoid the waveforms.}
\label{fig:erp}
\end{figure}

\begin{figure}[htbp]
\centering
\includegraphics[width=\textwidth]{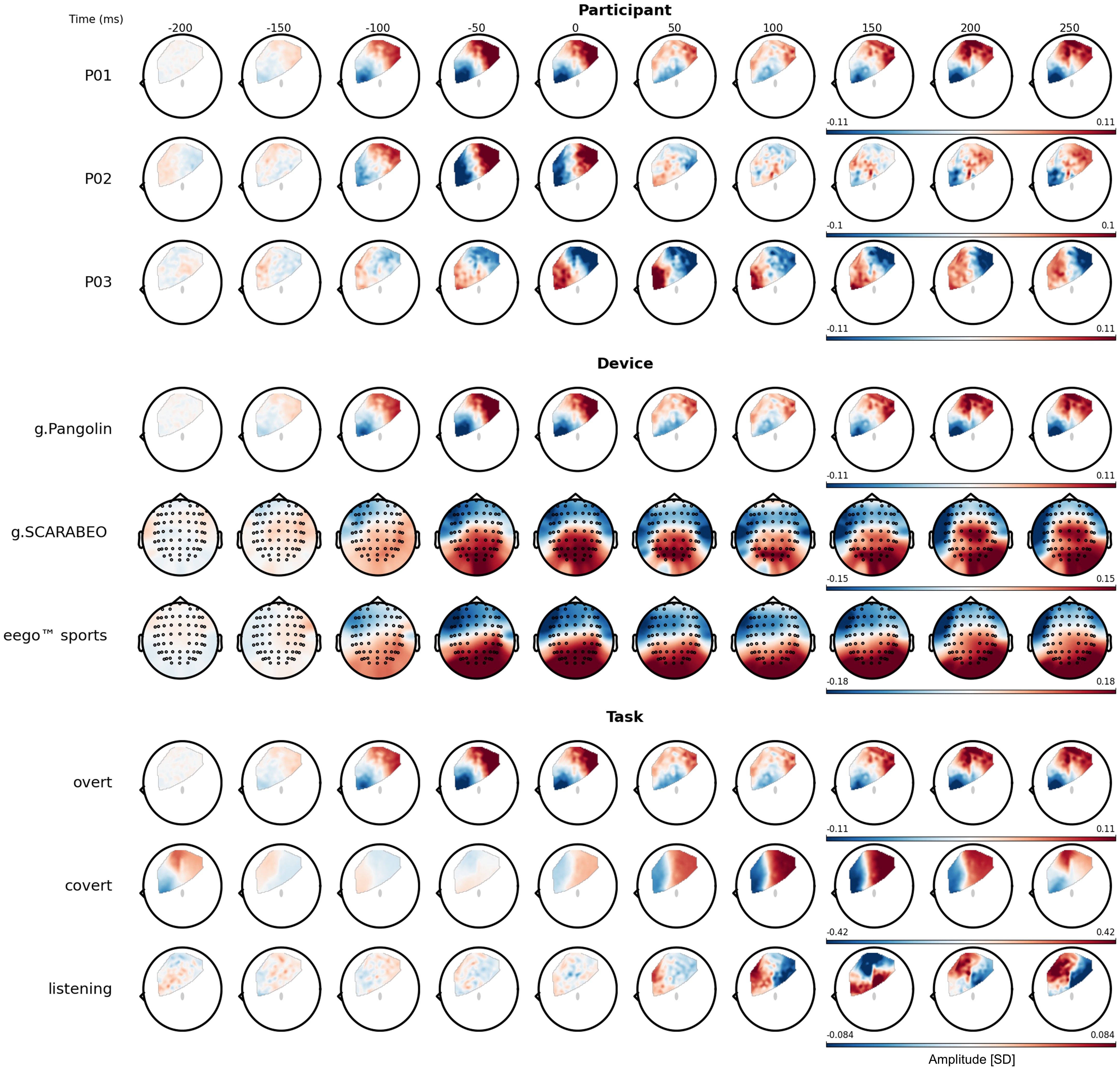}
\caption{Spatiotemporal dynamics of Event-Related Potentials (ERPs). The progression of spatial voltage distributions is shown from $-200$~ms to 250~ms after the event onset in 50~ms steps. Baselines and amplitude units are those of Figure~\ref{fig:erp}. Red and blue indicate positive and negative potentials, respectively. Panels display average distributions grouped by Participant (top row), Device (middle row), and Task (bottom row). Consistent with the validation strategy used in Figures~\ref{fig:psd} and~\ref{fig:erp}, specific parameters are held constant for each row: the Participant row shows data for the overt task using the g.Pangolin device; the Device row shows data for participant P01 during the overt task; and the Task row shows data for participant P01 using the g.Pangolin device. All maps are 2D scalp topographies interpolated between electrodes. The g.Pangolin grid covers the left lateral scalp. Its maps, which fill all Participant and Task rows, are therefore shown in a left lateral view with the nose to the left. They are restricted to the area covered by the electrodes. The g.SCARABEO and \eego{} maps are shown in the standard top view. The 0~ms mark denotes event onset. The maps show scalp potentials in sensor space. They are not cortical projections or source estimates.}
\label{fig:topo}
\end{figure}

\textbf{Event-related potentials.} To examine the temporal precision, spatial structure, and signal quality of the dataset, we analysed the event-related potentials (ERPs) and their topographical voltage distributions time-locked to the event onset (Figures~\ref{fig:erp} and~\ref{fig:topo}). We quantified the size of the evoked response relative to baseline fluctuations using the ratio of post-onset RMS amplitude to baseline standard deviation. For each channel, we averaged the ERPs across runs in which that channel was good. We calculated the resulting waveform's RMS amplitude from 0 to 400~ms after onset and its standard deviation during the baseline window. We averaged each measure across channels and divided the mean RMS by the mean baseline standard deviation. The baseline window was $-300$ to $-200$~ms for overt speech and $-200$ to 0~ms for covert speech and listening. This ratio ranged from 3.3 to 15.1 across participants, devices and conditions. It was lowest for listening (3.3), the condition with the fewest trials, and for covert speech (3.9). For P01 overt speech, the ratios were 13.0 for g.Pangolin, 12.1 for g.SCARABEO and 15.1 for \eego{}. The ERPs were also reproducible across runs. We divided the runs of each condition at random into two halves 200 times and correlated the two half averages over all channels from $-200$ to 800~ms. After the Spearman--Brown correction for the full number of runs, the reliability was 0.865 to 0.996 for overt and covert speech and 0.677 for listening. Following the event onset, the ERP traces diverge into positive and negative polarities with smooth amplitude gradients across spatially adjacent electrodes (Figure~\ref{fig:erp}). The scalp maps show spatially organized fields that evolve smoothly over time (Figure~\ref{fig:topo}). These maps show scalp potentials in sensor space and are not source estimates.

Responses time-locked to the event onset were present with all three EEG systems (Figure~\ref{fig:erp}, Device), with differences in their morphology. The g.Pangolin and \eego{} waveforms both showed an early and a later deflection. Relative to the speech-event onset, the GFP peaked at $-33$ and 208~ms for g.Pangolin and at $-29$ and 162~ms for \eego{}. The g.SCARABEO GFP peaked at $-4$ and 162~ms, and its scalp maps showed a front--back distribution similar to that of \eego{}, the other whole-head cap (Figure~\ref{fig:topo}). The panels show overt speech of P01 recorded in different periods with different utterances. The systems also differ in electrode coverage and acquisition reference (see EEG and EMG recording). The g.Pangolin grid covers the left lateral scalp, whereas g.SCARABEO and \eego{} are whole-head caps. All data were re-referenced to the good-channel average before analysis. The contribution of each factor to the observed differences cannot be determined from these recordings.

Inter-participant comparisons during the overt speech task showed both shared and individual response patterns (Figures~\ref{fig:erp} and~\ref{fig:topo}, Participant). P01 and P02 showed similar spatiotemporal progression and spatial response patterns, whereas P03 showed distinct spatial topographies and temporal profiles. P03 also spoke faster than the other participants. The articulation rate was computed from the released word and phone intervals. The morae of the aligned words of each utterance were counted and divided by the summed duration of their phones, so pauses were not counted. Utterances with words unknown to the aligner were excluded. The median rate was 7.8 morae per second (12.8 phones per second) for P01, 7.5 (12.4) for P02 and 11.3 (18.3) for P03. For comparison, the reported mean $\pm$ SD is 7.1 $\pm$ 1.0 morae per second for sentences read aloud by professional speakers and 8.0 $\pm$ 2.1 for utterances of spontaneous speech\cite{maekawa2003}. Relative to the spontaneous utterances, the medians of P01 and P02 lie near the mean, and that of P03 lies about 1.6 SD above it. The differences in the responses between participants are consistent with expected individual variation, such as differences in speech rate, articulatory habits, and head and electrode geometry.

Several features of the waveforms resemble known speech-related responses. During the continuous listening task, Voice Activity Detection (VAD)-aligned averaging yielded a multiphasic response with GFP peaks near 140 and 230~ms after the event onset (Figures~\ref{fig:erp} and~\ref{fig:topo}). The sound started about 40~ms after the event onset (median), so these peaks lie about 100 and 190~ms after the sound. These latencies are close to those of the auditory N1\cite{naatanen1987} and P2\cite{crowley2004}. This response is based on 19 runs (4,564 trials) of P01. It was reproduced from the released files with a correlation of 0.53 (Methods). Listening to speech engages auditory regions and can also evoke responses in motor cortex\cite{hickok2007,cheung2016}.

In overt speech, the ERP departed from the baseline about 150~ms before the speech-event onset (Figure~\ref{fig:erp}, Task: overt). It reached its first peak about 30~ms before the event onset, which is about 70~ms before the acoustic onset. This pre-onset deflection resembles the terminal phase of the readiness potential\cite{mcadam1971,brumberg2016}. Adaptive filtering was applied to suppress facial EMG, but the deflection may include residual muscle or movement-related contributions. In the covert condition, deflections were present within 300~ms after onset (Figure~\ref{fig:erp}, Task: covert). The g.Pangolin scalp maps for covert speech partly resembled those for overt speech (Figure~\ref{fig:topo}). Because each covert window was started with a key press, responses time-locked to its onset may also include activity related to the key press.

We assessed the influence of adaptive filtering on ERP morphology using P01 overt-speech data from 369 g.Pangolin, 137 g.SCARABEO and 172 \eego{} runs. The grand-average ERPs with and without adaptive filtering were highly similar. Their Pearson correlations over channel--time values from 0 to 800~ms were 0.9989, 0.9994 and 0.9997, respectively. RMS amplitude was 6.7\%, 6.2\% and 8.7\% lower with adaptive filtering. For the listening condition of P01 (19 g.Pangolin runs), the correlation was 0.987 and the RMS amplitude 3.5\% lower with adaptive filtering. Thus, the overall grand-average response was retained when adaptive filtering was omitted.

Together, the spectral profiles and time-locked responses observed with every system, including their amplitude relative to baseline fluctuations, indicate that the recordings are of suitable quality for reuse.

\section{Usage Notes}

The speech audio is voice-masked (see Voice masking). Its timing, linguistic content, and amplitude envelope follow the original recordings, but the participants' own voice timbre is not available. Analyses that require the original acoustic voice characteristics, such as reconstructing the participants' own voices from neural signals, therefore cannot be performed with the released audio. Event transcriptions were generated automatically and may contain recognition errors. For audiobook listening, the transcriptions identify the presented speech of each fragment. The order of the fragments and their position within the audiobooks are not provided.

Recording factors are partly confounded (Table~\ref{tab:events}). Comparisons between devices are possible only within P01. Each device was also used in a different period. We used g.SCARABEO from November 2023 to February 2024, \eego{} from August to November 2024, and g.Pangolin throughout. Device effects therefore cannot be separated from changes over time. The covert repetition task was recorded only for P01 with g.Pangolin, and P02 and P03 were recorded only with g.Pangolin. Comparisons between participants are therefore limited to overt speech with g.Pangolin, and comparisons between tasks to P01. When building speech-decoding models, nearby trials from the same recording can share temporal structure and session-specific signal characteristics. To evaluate generalisation to new sessions, we recommend splitting training and evaluation data by session or recording day. This is possible for the continuous recordings, which retain these dates. Results should be reported separately for each participant and device. Comparisons across participants, devices, tasks and sessions should be interpreted with care. The dataset is intended for within-participant speech decoding, cross-session and cross-device adaptation, artifact modeling, and EEG representation learning.

The raw recordings contain bad channels that differ between runs. In the continuous recordings, the median number of bad channels per run was three with g.Pangolin (interquartile range 1--6) and five with \eego{}. Of the g.Pangolin runs, 13\% had ten or more. A single broken channel included in a common average reference contaminates all other channels. We therefore recommend excluding bad channels before re-referencing. Bad channels of the continuous recordings are marked in the \texttt{status} column of the channel files, and the detection criteria are given in \texttt{status\_description}. Stretches in which the scalp EEG is saturated, or more than 1000 times its usual amplitude for the participant, device and task, are marked in the event files as \texttt{BAD\_high\_amplitude}. Stretches of 100 to 1000 times the usual amplitude are marked as \texttt{high\_amplitude}. The criteria are given in the README. Across all tasks, 0.6\% of trials overlap, at least in part, a stretch marked \texttt{BAD\_high\_amplitude}.

The validation analyses describe sensor-space signal characteristics. Speech production introduces muscular, ocular and movement-related contributions that may remain after preprocessing. Reference choice affects spectra and waveforms, and the observational comparisons do not isolate hardware effects or identify cortical generators. Utterances vary in content, duration and articulation. Averaging them at their onsets therefore emphasises consistently timed activity and may attenuate components with variable latencies. Users should choose artifact handling, references and baseline definitions appropriate for their analysis.

\section{Data Availability}

The JapanEEG dataset is available from Harvard Dataverse\cite{sato2026japaneeg} (\url{https://doi.org/10.7910/DVN/5RM3OT}) in Brain Imaging Data Structure (BIDS) format\cite{pernet2019}.
Access is controlled because the recordings contain voice-masked speech of the participants, readings of copyrighted texts, and transcriptions of commercial audiobooks. To obtain access, users log in to Harvard Dataverse, submit an access request stating their name, institutional affiliation, institutional e-mail address, and intended use, and accept the Data Use Agreement. Except for the corpus-derived materials described below, the agreement restricts use to non-commercial research involving the analysis of the data. The agreement also prohibits attempts to re-identify participants and redistribution of the data. Researchers may analyse the speech audio and transcriptions, including their linguistic content. The agreement prohibits using these materials primarily to enjoy, restore, or reconstruct the original works. Requests are reviewed by the authors, normally within two weeks. The speech audio and transcriptions include readings of copyrighted texts and the text of commercial audiobooks. They are provided for scholarly information analysis, principally feature extraction and the training and evaluation of models that relate neural signals to speech. This provision relies on the copyright limitations for information analysis (Article 30-4 of the Japanese Copyright Act) and on fair use or fair dealing where applicable. Users are responsible for complying with the laws that apply to them. Utterance-level segments of these readings are provided in random order, without information on their original order. The recordings of the covert repetition task and of audiobook listening are provided in the same way. The replayed segments of the former can include such readings, and the transcriptions of the latter contain the text of the works.

The speech audio and transcriptions of the JSUT and ReazonSpeech readings are identified separately in Annex A of the Data Use Agreement. For JSUT-derived materials, the original text licence notices (CC BY-SA 3.0, CC BY 2.0 and CC BY-SA 4.0, as applicable) are retained. Rights held by the Provider in the newly created speech recordings are made available under CC BY-SA 4.0. ReazonSpeech-derived speech audio and transcriptions are provided under CDLA-Sharing-1.0 with the corpus's information-analysis purpose limitation. For these identified audio and transcription files, the original terms take precedence over the general DUA restrictions, subject to participant and personal-data protections that do not conflict with those terms. EEG, EMG and all other components remain subject to the DUA.

\section{Code Availability}

The JapanEEG code repository is available at \url{https://github.com/Motoshige496/JapanEEG}.

\section{Author Contributions}

M.S. (Motoshige Sato): Conceptualization, Data curation, Formal analysis, Methodology, Project administration, Software, Validation, Visualization, Writing -- review \& editing.
I.H. (Ilya Horiguchi): Data curation, Software, Resources, Project administration, Validation.
M.I. (Masakazu Inoue): Conceptualization, Data curation, Investigation, Validation.
K.T. (Kenichi Tomeoka): Data curation, Project administration, Software.
E.H. (Eri Hatakeyama): Data curation, Project administration, Software.
Y.K. (Yuya Kita): Data curation, Project administration.
A.Y. (Atsushi Yamamoto): Data curation, Project administration.
I.F. (Ippei Fujisawa): Conceptualization, Methodology.
S.S. (Shuntaro Sasai): Conceptualization, Funding acquisition, Investigation, Methodology, Project administration, Resources, Supervision, Visualization, Writing -- original draft, Writing -- review \& editing.

\section{Competing Interests}

The authors declare no competing interests.

\section{Funding}

This work was supported by JST, Moonshot R\&D Grant Number JPMJMS2012.

\section{Acknowledgements}

We thank Yasuo Kabe, Sensho Nobe, and Akito Yoshida for valuable discussions on the research direction and data collection strategy during the early stages of this project. We are grateful to Mayumi Shimizu for her support in data collection and administrative coordination, and to Ryu Miyata for administrative support. We thank Yukihito Yomogida for his continued administrative support, including the preparation of ethics committee documentation and other regulatory procedures essential to this study. We also thank Ryota Kanai for his guidance as program manager of the grant program and for discussions on the data collection strategy, and Kai Arulkumaran for helpful discussions on the data collection approach.

\bibliographystyle{naturemag}
\bibliography{references}

\begin{thebibliography}{10}
\expandafter\ifx\csname url\endcsname\relax
  \def\url#1{\texttt{#1}}\fi
\expandafter\ifx\csname urlprefix\endcsname\relax\def\urlprefix{URL }\fi
\providecommand{\bibinfo}[2]{#2}
\providecommand{\eprint}[2][]{\url{#2}}

\bibitem{tudor2005}
\bibinfo{author}{Tudor, M.}, \bibinfo{author}{Tudor, L.} \&
  \bibinfo{author}{Tudor, K.~I.}
\newblock \bibinfo{title}{{Hans Berger} (1873--1941)---the history of
  electroencephalography}.
\newblock \emph{\bibinfo{journal}{Acta Med. Croatica}}
  \textbf{\bibinfo{volume}{59}}, \bibinfo{pages}{307--313}
  (\bibinfo{year}{2005}).
\newblock \urlprefix\url{https://pubmed.ncbi.nlm.nih.gov/16334737/}.

\bibitem{moses2021}
\bibinfo{author}{Moses, D.~A.} \emph{et~al.}
\newblock \bibinfo{title}{Neuroprosthesis for decoding speech in a paralyzed
  person with anarthria}.
\newblock \emph{\bibinfo{journal}{N. Engl. J. Med.}}
  \textbf{\bibinfo{volume}{385}}, \bibinfo{pages}{217--227}
  (\bibinfo{year}{2021}).
\newblock \urlprefix\url{https://doi.org/10.1056/nejmoa2027540}.

\bibitem{willett2023}
\bibinfo{author}{Willett, F.~R.} \emph{et~al.}
\newblock \bibinfo{title}{A high-performance speech neuroprosthesis}.
\newblock \emph{\bibinfo{journal}{Nature}} \textbf{\bibinfo{volume}{620}},
  \bibinfo{pages}{1031--1036} (\bibinfo{year}{2023}).
\newblock \urlprefix\url{https://doi.org/10.1038/s41586-023-06377-x}.

\bibitem{metzger2023}
\bibinfo{author}{Metzger, S.~L.} \emph{et~al.}
\newblock \bibinfo{title}{A high-performance neuroprosthesis for speech
  decoding and avatar control}.
\newblock \emph{\bibinfo{journal}{Nature}} \textbf{\bibinfo{volume}{620}},
  \bibinfo{pages}{1037--1046} (\bibinfo{year}{2023}).
\newblock \urlprefix\url{https://doi.org/10.1038/s41586-023-06443-4}.

\bibitem{card2024}
\bibinfo{author}{Card, N.~S.} \emph{et~al.}
\newblock \bibinfo{title}{An accurate and rapidly calibrating speech
  neuroprosthesis}.
\newblock \emph{\bibinfo{journal}{N. Engl. J. Med.}}
  \textbf{\bibinfo{volume}{391}}, \bibinfo{pages}{609--618}
  (\bibinfo{year}{2024}).
\newblock \urlprefix\url{https://doi.org/10.1056/nejmoa2314132}.

\bibitem{anumanchipalli2019}
\bibinfo{author}{Anumanchipalli, G.~K.}, \bibinfo{author}{Chartier, J.} \&
  \bibinfo{author}{Chang, E.~F.}
\newblock \bibinfo{title}{Speech synthesis from neural decoding of spoken
  sentences}.
\newblock \emph{\bibinfo{journal}{Nature}} \textbf{\bibinfo{volume}{568}},
  \bibinfo{pages}{493--498} (\bibinfo{year}{2019}).
\newblock \urlprefix\url{https://doi.org/10.1038/s41586-019-1119-1}.

\bibitem{proix2022}
\bibinfo{author}{Proix, T.} \emph{et~al.}
\newblock \bibinfo{title}{Imagined speech can be decoded from low- and
  cross-frequency intracranial {EEG} features}.
\newblock \emph{\bibinfo{journal}{Nat. Commun.}} \textbf{\bibinfo{volume}{13}},
  \bibinfo{pages}{48} (\bibinfo{year}{2022}).
\newblock \urlprefix\url{https://doi.org/10.1038/s41467-021-27725-3}.

\bibitem{dascoli2025}
\bibinfo{author}{d'Ascoli, S.} \emph{et~al.}
\newblock \bibinfo{title}{Towards decoding individual words from non-invasive
  brain recordings}.
\newblock \emph{\bibinfo{journal}{Nat. Commun.}} \textbf{\bibinfo{volume}{16}},
  \bibinfo{pages}{10521} (\bibinfo{year}{2025}).
\newblock \urlprefix\url{https://doi.org/10.1038/s41467-025-65499-0}.

\bibitem{suppes1997}
\bibinfo{author}{Suppes, P.}, \bibinfo{author}{Lu, Z.-L.} \&
  \bibinfo{author}{Han, B.}
\newblock \bibinfo{title}{Brain wave recognition of words}.
\newblock \emph{\bibinfo{journal}{Proc. Natl Acad. Sci. USA}}
  \textbf{\bibinfo{volume}{94}}, \bibinfo{pages}{14965--14969}
  (\bibinfo{year}{1997}).
\newblock \urlprefix\url{https://doi.org/10.1073/pnas.94.26.14965}.

\bibitem{lee2023}
\bibinfo{author}{Lee, Y.-E.}, \bibinfo{author}{Lee, S.-H.},
  \bibinfo{author}{Kim, S.-H.} \& \bibinfo{author}{Lee, S.-W.}
\newblock \bibinfo{title}{Towards voice reconstruction from {EEG} during
  imagined speech}.
\newblock \emph{\bibinfo{journal}{Proc. AAAI Conf. Artif. Intell.}}
  \textbf{\bibinfo{volume}{37}}, \bibinfo{pages}{6030--6038}
  (\bibinfo{year}{2023}).
\newblock \urlprefix\url{https://doi.org/10.1609/aaai.v37i5.25745}.

\bibitem{kaongoen2022}
\bibinfo{author}{Kaongoen, N.}, \bibinfo{author}{Choi, J.} \&
  \bibinfo{author}{Jo, S.}
\newblock \bibinfo{title}{A novel online {BCI} system using speech imagery and
  ear-{EEG} for home appliances control}.
\newblock \emph{\bibinfo{journal}{Comput. Methods Programs Biomed.}}
  \textbf{\bibinfo{volume}{224}}, \bibinfo{pages}{107022}
  (\bibinfo{year}{2022}).
\newblock \urlprefix\url{https://doi.org/10.1016/j.cmpb.2022.107022}.

\bibitem{lopezbernal2022}
\bibinfo{author}{Lopez-Bernal, D.}, \bibinfo{author}{Balderas, D.},
  \bibinfo{author}{Ponce, P.} \& \bibinfo{author}{Molina, A.}
\newblock \bibinfo{title}{A state-of-the-art review of {EEG}-based imagined
  speech decoding}.
\newblock \emph{\bibinfo{journal}{Front. Hum. Neurosci.}}
  \textbf{\bibinfo{volume}{16}}, \bibinfo{pages}{867281}
  (\bibinfo{year}{2022}).
\newblock \urlprefix\url{https://doi.org/10.3389/fnhum.2022.867281}.

\bibitem{panachakel2021}
\bibinfo{author}{Panachakel, J.~T.} \& \bibinfo{author}{Ramakrishnan, A.~G.}
\newblock \bibinfo{title}{Decoding covert speech from {EEG}---a comprehensive
  review}.
\newblock \emph{\bibinfo{journal}{Front. Neurosci.}}
  \textbf{\bibinfo{volume}{15}}, \bibinfo{pages}{642251}
  (\bibinfo{year}{2021}).
\newblock \urlprefix\url{https://doi.org/10.3389/fnins.2021.642251}.

\bibitem{kaplan2020}
\bibinfo{author}{Kaplan, J.} \emph{et~al.}
\newblock \bibinfo{title}{Scaling laws for neural language models}.
\newblock \bibinfo{howpublished}{Preprint at
  \url{https://arxiv.org/abs/2001.08361}} (\bibinfo{year}{2020}).

\bibitem{banville2025}
\bibinfo{author}{Banville, H.}, \bibinfo{author}{Benchetrit, Y.},
  \bibinfo{author}{d'Ascoli, S.}, \bibinfo{author}{Rapin, J.} \&
  \bibinfo{author}{King, J.-R.}
\newblock \bibinfo{title}{Scaling laws for decoding images from brain
  activity}.
\newblock \bibinfo{howpublished}{Preprint at
  \url{https://arxiv.org/abs/2501.15322}} (\bibinfo{year}{2025}).

\bibitem{obeid2016}
\bibinfo{author}{Obeid, I.} \& \bibinfo{author}{Picone, J.}
\newblock \bibinfo{title}{The {Temple University Hospital EEG Data Corpus}}.
\newblock \emph{\bibinfo{journal}{Front. Neurosci.}}
  \textbf{\bibinfo{volume}{10}}, \bibinfo{pages}{196} (\bibinfo{year}{2016}).
\newblock \urlprefix\url{https://doi.org/10.3389/fnins.2016.00196}.

\bibitem{grootswagers2022}
\bibinfo{author}{Grootswagers, T.}, \bibinfo{author}{Zhou, I.},
  \bibinfo{author}{Robinson, A.~K.}, \bibinfo{author}{Hebart, M.~N.} \&
  \bibinfo{author}{Carlson, T.~A.}
\newblock \bibinfo{title}{Human {EEG} recordings for 1,854 concepts presented
  in rapid serial visual presentation streams}.
\newblock \emph{\bibinfo{journal}{Sci. Data}} \textbf{\bibinfo{volume}{9}},
  \bibinfo{pages}{3} (\bibinfo{year}{2022}).
\newblock \urlprefix\url{https://doi.org/10.1038/s41597-021-01102-7}.

\bibitem{defossez2023}
\bibinfo{author}{D{\'e}fossez, A.}, \bibinfo{author}{Caucheteux, C.},
  \bibinfo{author}{Rapin, J.}, \bibinfo{author}{Kabeli, O.} \&
  \bibinfo{author}{King, J.-R.}
\newblock \bibinfo{title}{Decoding speech perception from non-invasive brain
  recordings}.
\newblock \emph{\bibinfo{journal}{Nat. Mach. Intell.}}
  \textbf{\bibinfo{volume}{5}}, \bibinfo{pages}{1097--1107}
  (\bibinfo{year}{2023}).
\newblock \urlprefix\url{https://doi.org/10.1038/s42256-023-00714-5}.

\bibitem{schoffelen2019}
\bibinfo{author}{Schoffelen, J.-M.} \emph{et~al.}
\newblock \bibinfo{title}{A 204-subject multimodal neuroimaging dataset to
  study language processing}.
\newblock \emph{\bibinfo{journal}{Sci. Data}} \textbf{\bibinfo{volume}{6}},
  \bibinfo{pages}{17} (\bibinfo{year}{2019}).
\newblock \urlprefix\url{https://doi.org/10.1038/s41597-019-0020-y}.

\bibitem{gwilliams2023}
\bibinfo{author}{Gwilliams, L.} \emph{et~al.}
\newblock \bibinfo{title}{Introducing {MEG-MASC} a high-quality
  magneto-encephalography dataset for evaluating natural speech processing}.
\newblock \emph{\bibinfo{journal}{Sci. Data}} \textbf{\bibinfo{volume}{10}},
  \bibinfo{pages}{862} (\bibinfo{year}{2023}).
\newblock \urlprefix\url{https://doi.org/10.1038/s41597-023-02752-5}.

\bibitem{kostas2021}
\bibinfo{author}{Kostas, D.}, \bibinfo{author}{Aroca-Ouellette, S.} \&
  \bibinfo{author}{Rudzicz, F.}
\newblock \bibinfo{title}{{BENDR}: using transformers and a contrastive
  self-supervised learning task to learn from massive amounts of {EEG} data}.
\newblock \emph{\bibinfo{journal}{Front. Hum. Neurosci.}}
  \textbf{\bibinfo{volume}{15}}, \bibinfo{pages}{653659}
  (\bibinfo{year}{2021}).
\newblock \urlprefix\url{https://doi.org/10.3389/fnhum.2021.653659}.

\bibitem{zhu2023eeg2vec}
\bibinfo{author}{Zhu, Q.} \emph{et~al.}
\newblock \bibinfo{title}{{Eeg2vec}: self-supervised electroencephalographic
  representation learning}.
\newblock \bibinfo{howpublished}{Preprint at
  \url{https://arxiv.org/abs/2305.13957}} (\bibinfo{year}{2023}).

\bibitem{jiang2024}
\bibinfo{author}{Jiang, W.-B.}, \bibinfo{author}{Zhao, L.-M.} \&
  \bibinfo{author}{Lu, B.-L.}
\newblock \bibinfo{title}{Large brain model for learning generic
  representations with tremendous {EEG} data in {BCI}}.
\newblock In \emph{\bibinfo{booktitle}{The Twelfth International Conference on
  Learning Representations}} (\bibinfo{year}{2024}).
\newblock \urlprefix\url{https://arxiv.org/abs/2405.18765}.

\bibitem{wang2025}
\bibinfo{author}{Wang, J.} \emph{et~al.}
\newblock \bibinfo{title}{{CBraMod}: a criss-cross brain foundation model for
  {EEG} decoding}.
\newblock In \emph{\bibinfo{booktitle}{The Thirteenth International Conference
  on Learning Representations}} (\bibinfo{year}{2025}).
\newblock \urlprefix\url{https://arxiv.org/abs/2412.07236}.

\bibitem{yue2024}
\bibinfo{author}{Yue, T.} \emph{et~al.}
\newblock \bibinfo{title}{{EEGPT}: unleashing the potential of {EEG} generalist
  foundation model by autoregressive pre-training}.
\newblock \bibinfo{howpublished}{Preprint at
  \url{https://arxiv.org/abs/2410.19779v1}} (\bibinfo{year}{2024}).

\bibitem{elouahidi2025}
\bibinfo{author}{El~Ouahidi, Y.} \emph{et~al.}
\newblock \bibinfo{title}{{REVE}: a foundation model for {EEG}---adapting to
  any setup with large-scale pretraining on 25,000 subjects}.
\newblock In \emph{\bibinfo{booktitle}{Advances in Neural Information
  Processing Systems}}, vol.~\bibinfo{volume}{38},
  \bibinfo{pages}{25583--25619} (\bibinfo{year}{2025}).
\newblock \urlprefix\url{https://doi.org/10.52202/085713-0760}.

\bibitem{saha2020}
\bibinfo{author}{Saha, S.} \& \bibinfo{author}{Baumert, M.}
\newblock \bibinfo{title}{Intra- and inter-subject variability in {EEG}-based
  sensorimotor brain computer interface: a review}.
\newblock \emph{\bibinfo{journal}{Front. Comput. Neurosci.}}
  \textbf{\bibinfo{volume}{13}}, \bibinfo{pages}{87} (\bibinfo{year}{2020}).
\newblock \urlprefix\url{https://doi.org/10.3389/fncom.2019.00087}.

\bibitem{kuruppu2025}
\bibinfo{author}{Kuruppu, G.}, \bibinfo{author}{Wagh, N.},
  \bibinfo{author}{Kremen, V.} \& \bibinfo{author}{Varatharajah, Y.}
\newblock \bibinfo{title}{{EEG} foundation models: a critical review of current
  progress and future directions}.
\newblock \emph{\bibinfo{journal}{J. Neural Eng.}}
  \textbf{\bibinfo{volume}{23}}, \bibinfo{pages}{021001}
  (\bibinfo{year}{2026}).
\newblock \urlprefix\url{https://doi.org/10.1088/1741-2552/ae4455}.

\bibitem{heo2025}
\bibinfo{author}{Heo, D.} \& \bibinfo{author}{Kim, S.-P.}
\newblock \bibinfo{title}{Freeing {P300}-based brain-computer interfaces from
  daily recalibration by extracting daily common {ERPs}}.
\newblock \emph{\bibinfo{journal}{IEEE Trans. Neural Syst. Rehabil. Eng.}}
  \textbf{\bibinfo{volume}{33}}, \bibinfo{pages}{2977--2987}
  (\bibinfo{year}{2025}).
\newblock \urlprefix\url{https://doi.org/10.1109/tnsre.2025.3594341}.

\bibitem{huang2021}
\bibinfo{author}{Huang, X.} \emph{et~al.}
\newblock \bibinfo{title}{A review on signal processing approaches to reduce
  calibration time in {EEG}-based brain--computer interface}.
\newblock \emph{\bibinfo{journal}{Front. Neurosci.}}
  \textbf{\bibinfo{volume}{15}}, \bibinfo{pages}{733546}
  (\bibinfo{year}{2021}).
\newblock \urlprefix\url{https://doi.org/10.3389/fnins.2021.733546}.

\bibitem{kamrud2021}
\bibinfo{author}{Kamrud, A.}, \bibinfo{author}{Borghetti, B.} \&
  \bibinfo{author}{Schubert~Kabban, C.}
\newblock \bibinfo{title}{The effects of individual differences,
  non-stationarity, and the importance of data partitioning decisions for
  training and testing of {EEG} cross-participant models}.
\newblock \emph{\bibinfo{journal}{Sensors}} \textbf{\bibinfo{volume}{21}},
  \bibinfo{pages}{3225} (\bibinfo{year}{2021}).
\newblock \urlprefix\url{https://doi.org/10.3390/s21093225}.

\bibitem{peterson2021}
\bibinfo{author}{Peterson, S.~M.}, \bibinfo{author}{Steine-Hanson, Z.},
  \bibinfo{author}{Davis, N.}, \bibinfo{author}{Rao, R. P.~N.} \&
  \bibinfo{author}{Brunton, B.~W.}
\newblock \bibinfo{title}{Generalized neural decoders for transfer learning
  across participants and recording modalities}.
\newblock \emph{\bibinfo{journal}{J. Neural Eng.}}
  \textbf{\bibinfo{volume}{18}}, \bibinfo{pages}{026014}
  (\bibinfo{year}{2021}).
\newblock \urlprefix\url{https://doi.org/10.1088/1741-2552/abda0b}.

\bibitem{zhao2015}
\bibinfo{author}{Zhao, S.} \& \bibinfo{author}{Rudzicz, F.}
\newblock \bibinfo{title}{Classifying phonological categories in imagined and
  articulated speech}.
\newblock In \emph{\bibinfo{booktitle}{2015 IEEE International Conference on
  Acoustics, Speech and Signal Processing (ICASSP)}}, \bibinfo{pages}{992--996}
  (\bibinfo{publisher}{IEEE}, \bibinfo{year}{2015}).
\newblock \urlprefix\url{https://doi.org/10.1109/icassp.2015.7178118}.

\bibitem{dekker2023}
\bibinfo{author}{Dekker, B.}, \bibinfo{author}{Schouten, A.~C.} \&
  \bibinfo{author}{Scharenborg, O.}
\newblock \bibinfo{title}{{DAIS}: the {Delft} database of {EEG} recordings of
  {Dutch} articulated and imagined speech}.
\newblock In \emph{\bibinfo{booktitle}{ICASSP 2023 -- 2023 IEEE International
  Conference on Acoustics, Speech and Signal Processing (ICASSP)}},
  \bibinfo{pages}{1--5} (\bibinfo{publisher}{IEEE}, \bibinfo{year}{2023}).
\newblock \urlprefix\url{https://doi.org/10.1109/icassp49357.2023.10096145}.

\bibitem{zhang2024}
\bibinfo{author}{Zhang, Z.} \emph{et~al.}
\newblock \bibinfo{title}{Chisco: an {EEG}-based {BCI} dataset for decoding of
  imagined speech}.
\newblock \emph{\bibinfo{journal}{Sci. Data}} \textbf{\bibinfo{volume}{11}},
  \bibinfo{pages}{1265} (\bibinfo{year}{2024}).
\newblock \urlprefix\url{https://doi.org/10.1038/s41597-024-04114-1}.

\bibitem{moreira2025}
\bibinfo{author}{Moreira, J. P.~C.} \emph{et~al.}
\newblock \bibinfo{title}{An open-access {EEG} dataset for speech decoding:
  exploring the role of articulation and coarticulation}.
\newblock \emph{\bibinfo{journal}{Sci. Data}} \textbf{\bibinfo{volume}{12}},
  \bibinfo{pages}{1017} (\bibinfo{year}{2025}).
\newblock \urlprefix\url{https://doi.org/10.1038/s41597-025-05187-2}.

\bibitem{tan2025}
\bibinfo{author}{Tan, C.} \& \bibinfo{author}{Zhang, Q.}
\newblock \bibinfo{title}{Open access dataset integrating behavioral and {EEG}
  measures in {Chinese} spoken word production}.
\newblock \emph{\bibinfo{journal}{Sci. Data}} \textbf{\bibinfo{volume}{12}},
  \bibinfo{pages}{1348} (\bibinfo{year}{2025}).
\newblock \urlprefix\url{https://doi.org/10.1038/s41597-025-05671-9}.

\bibitem{zhao2025}
\bibinfo{author}{Zhao, R.} \emph{et~al.}
\newblock \bibinfo{title}{An open dataset of multidimensional signals based on
  different speech patterns in pragmatic {Mandarin}}.
\newblock \emph{\bibinfo{journal}{Sci. Data}} \textbf{\bibinfo{volume}{12}},
  \bibinfo{pages}{1934} (\bibinfo{year}{2025}).
\newblock \urlprefix\url{https://doi.org/10.1038/s41597-025-06213-z}.

\bibitem{chen2026}
\bibinfo{author}{Chen, S.} \emph{et~al.}
\newblock \bibinfo{title}{An {EEG} dataset for multimodal semantic alignment
  and neural decoding during reading and listening}.
\newblock \emph{\bibinfo{journal}{Sci. Data}} \textbf{\bibinfo{volume}{13}},
  \bibinfo{pages}{148} (\bibinfo{year}{2026}).
\newblock \urlprefix\url{https://doi.org/10.1038/s41597-025-06466-8}.

\bibitem{nieto2022}
\bibinfo{author}{Nieto, N.}, \bibinfo{author}{Peterson, V.},
  \bibinfo{author}{Rufiner, H.~L.}, \bibinfo{author}{Kamienkowski, J.~E.} \&
  \bibinfo{author}{Spies, R.}
\newblock \bibinfo{title}{Thinking out loud, an open-access {EEG}-based {BCI}
  dataset for inner speech recognition}.
\newblock \emph{\bibinfo{journal}{Sci. Data}} \textbf{\bibinfo{volume}{9}},
  \bibinfo{pages}{52} (\bibinfo{year}{2022}).
\newblock \urlprefix\url{https://doi.org/10.1038/s41597-022-01147-2}.

\bibitem{craik2025}
\bibinfo{author}{Craik, A.}, \bibinfo{author}{Dial, H.} \&
  \bibinfo{author}{Contreras-Vidal, J.~L.}
\newblock \bibinfo{title}{Continuous and discrete decoding of overt speech with
  scalp electroencephalography ({EEG})}.
\newblock \emph{\bibinfo{journal}{J. Neural Eng.}}
  \textbf{\bibinfo{volume}{22}}, \bibinfo{pages}{026017}
  (\bibinfo{year}{2025}).
\newblock \urlprefix\url{https://doi.org/10.1088/1741-2552/ad8d0a}.

\bibitem{accou2024}
\bibinfo{author}{Accou, B.} \emph{et~al.}
\newblock \bibinfo{title}{{SparrKULee}: a speech-evoked auditory response
  repository from {KU} {L}euven, containing the {EEG} of 85 participants}.
\newblock \emph{\bibinfo{journal}{Data}} \textbf{\bibinfo{volume}{9}},
  \bibinfo{pages}{94} (\bibinfo{year}{2024}).
\newblock \urlprefix\url{https://doi.org/10.3390/data9080094}.

\bibitem{xu2024}
\bibinfo{author}{Xu, X.} \emph{et~al.}
\newblock \bibinfo{title}{{ConvConcatNet}: a deep convolutional neural network
  to reconstruct mel spectrogram from the {EEG}}.
\newblock In \emph{\bibinfo{booktitle}{2024 IEEE International Conference on
  Acoustics, Speech, and Signal Processing Workshops (ICASSPW)}},
  \bibinfo{pages}{113--114} (\bibinfo{year}{2024}).
\newblock \urlprefix\url{https://doi.org/10.1109/icasspw62465.2024.10626859}.

\bibitem{pei2011}
\bibinfo{author}{Pei, X.} \emph{et~al.}
\newblock \bibinfo{title}{Spatiotemporal dynamics of electrocorticographic high
  gamma activity during overt and covert word repetition}.
\newblock \emph{\bibinfo{journal}{NeuroImage}} \textbf{\bibinfo{volume}{54}},
  \bibinfo{pages}{2960--2972} (\bibinfo{year}{2011}).
\newblock \urlprefix\url{https://doi.org/10.1016/j.neuroimage.2010.10.029}.

\bibitem{brumberg2016}
\bibinfo{author}{Brumberg, J.~S.} \emph{et~al.}
\newblock \bibinfo{title}{Spatio-temporal progression of cortical activity
  related to continuous overt and covert speech production in a reading task}.
\newblock \emph{\bibinfo{journal}{PLoS ONE}} \textbf{\bibinfo{volume}{11}},
  \bibinfo{pages}{e0166872} (\bibinfo{year}{2016}).
\newblock \urlprefix\url{https://doi.org/10.1371/journal.pone.0166872}.

\bibitem{komeiji2024}
\bibinfo{author}{Komeiji, S.} \emph{et~al.}
\newblock \bibinfo{title}{Feasibility of decoding covert speech in {ECoG} with
  a {Transformer} trained on overt speech}.
\newblock \emph{\bibinfo{journal}{Sci. Rep.}} \textbf{\bibinfo{volume}{14}},
  \bibinfo{pages}{11491} (\bibinfo{year}{2024}).
\newblock \urlprefix\url{https://doi.org/10.1038/s41598-024-62230-9}.

\bibitem{kunz2025}
\bibinfo{author}{Kunz, E.~M.} \emph{et~al.}
\newblock \bibinfo{title}{Inner speech in motor cortex and implications for
  speech neuroprostheses}.
\newblock \emph{\bibinfo{journal}{Cell}} \textbf{\bibinfo{volume}{188}},
  \bibinfo{pages}{4658--4673.e17} (\bibinfo{year}{2025}).
\newblock \urlprefix\url{https://doi.org/10.1016/j.cell.2025.06.015}.

\bibitem{stephan2020}
\bibinfo{author}{Stephan, F.}, \bibinfo{author}{Saalbach, H.} \&
  \bibinfo{author}{Rossi, S.}
\newblock \bibinfo{title}{The brain differentially prepares inner and overt
  speech production: electrophysiological and vascular evidence}.
\newblock \emph{\bibinfo{journal}{Brain Sci.}} \textbf{\bibinfo{volume}{10}},
  \bibinfo{pages}{148} (\bibinfo{year}{2020}).
\newblock \urlprefix\url{https://doi.org/10.3390/brainsci10030148}.

\bibitem{devos2010}
\bibinfo{author}{De~Vos, M.} \emph{et~al.}
\newblock \bibinfo{title}{Removal of muscle artifacts from {EEG} recordings of
  spoken language production}.
\newblock \emph{\bibinfo{journal}{Neuroinformatics}}
  \textbf{\bibinfo{volume}{8}}, \bibinfo{pages}{135--150}
  (\bibinfo{year}{2010}).
\newblock \urlprefix\url{https://doi.org/10.1007/s12021-010-9071-0}.

\bibitem{porcaro2015}
\bibinfo{author}{Porcaro, C.}, \bibinfo{author}{Medaglia, M.~T.} \&
  \bibinfo{author}{Krott, A.}
\newblock \bibinfo{title}{Removing speech artifacts from
  electroencephalographic recordings during overt picture naming}.
\newblock \emph{\bibinfo{journal}{NeuroImage}} \textbf{\bibinfo{volume}{105}},
  \bibinfo{pages}{171--180} (\bibinfo{year}{2015}).
\newblock \urlprefix\url{https://doi.org/10.1016/j.neuroimage.2014.10.049}.

\bibitem{sato2026jne}
\bibinfo{author}{Sato, M.} \emph{et~al.}
\newblock \bibinfo{title}{Delineating neural contributions to
  electroencephalogram-based speech decoding}.
\newblock \emph{\bibinfo{journal}{J. Neural Eng.}}
  \textbf{\bibinfo{volume}{23}}, \bibinfo{pages}{056062}
  (\bibinfo{year}{2026}).
\newblock \urlprefix\url{https://doi.org/10.1088/1741-2552/aeb68d}.

\bibitem{sato2024}
\bibinfo{author}{Sato, M.} \emph{et~al.}
\newblock \bibinfo{title}{Scaling law in neural data: non-invasive speech
  decoding with 175 hours of {EEG} data}.
\newblock \bibinfo{howpublished}{Preprint at
  \url{https://arxiv.org/abs/2407.07595}} (\bibinfo{year}{2024}).

\bibitem{sato2026large}
\bibinfo{author}{Sato, M.} \emph{et~al.}
\newblock \bibinfo{title}{Large-scale data practicalizes {EEG}-based speech
  decoding}.
\newblock In \bibinfo{editor}{Guger, C.}, \bibinfo{editor}{Allison, B.},
  \bibinfo{editor}{Lotte, F.} \& \bibinfo{editor}{Korostenskaja, M.} (eds.)
  \emph{\bibinfo{booktitle}{Brain-Computer Interface Research: A
  State-of-the-Art Summary 13}}, SpringerBriefs in Electrical and Computer
  Engineering, \bibinfo{pages}{123--129} (\bibinfo{publisher}{Springer Nature
  Switzerland}, \bibinfo{address}{Cham}, \bibinfo{year}{2026}).
\newblock \urlprefix\url{https://doi.org/10.1007/978-3-032-25828-1_13}.

\bibitem{sato2026japaneeg}
\bibinfo{author}{Sato, M.}, \bibinfo{author}{Horiguchi, I.},
  \bibinfo{author}{Inoue, M.} \emph{et~al.}
\newblock \bibinfo{title}{{JapanEEG}: a 1000-hour {EEG}--{EMG}--audio dataset
  of {Japanese} speech production}.
\newblock \bibinfo{howpublished}{Harvard Dataverse} (\bibinfo{year}{2026}).
\newblock \bibinfo{note}{\url{https://doi.org/10.7910/DVN/5RM3OT}}.

\bibitem{silero2024}
\bibinfo{author}{{Silero Team}}.
\newblock \bibinfo{title}{{Silero VAD}: pre-trained enterprise-grade voice
  activity detector ({VAD}), number detector and language classifier}.
\newblock \bibinfo{howpublished}{GitHub repository} (\bibinfo{year}{2024}).
\newblock \urlprefix\url{https://github.com/snakers4/silero-vad}.

\bibitem{desplanques2020}
\bibinfo{author}{Desplanques, B.}, \bibinfo{author}{Thienpondt, J.} \&
  \bibinfo{author}{Demuynck, K.}
\newblock \bibinfo{title}{{ECAPA-TDNN}: Emphasized channel attention,
  propagation and aggregation in {TDNN} based speaker verification}.
\newblock In \emph{\bibinfo{booktitle}{Proc. Interspeech}},
  \bibinfo{pages}{3830--3834} (\bibinfo{year}{2020}).
\newblock \urlprefix\url{https://doi.org/10.21437/interspeech.2020-2650}.

\bibitem{seedvc2024}
\bibinfo{author}{Liu, S.}
\newblock \bibinfo{title}{Zero-shot voice conversion with diffusion
  transformers}.
\newblock \bibinfo{howpublished}{Preprint at
  \url{https://arxiv.org/abs/2411.09943}} (\bibinfo{year}{2024}).

\bibitem{radford2023whisper}
\bibinfo{author}{Radford, A.} \emph{et~al.}
\newblock \bibinfo{title}{Robust speech recognition via large-scale weak
  supervision}.
\newblock In \emph{\bibinfo{booktitle}{Proceedings of the 40th International
  Conference on Machine Learning}}, vol. \bibinfo{volume}{202} of
  \emph{\bibinfo{series}{Proceedings of Machine Learning Research}},
  \bibinfo{pages}{28492--28518} (\bibinfo{year}{2023}).
\newblock \urlprefix\url{https://proceedings.mlr.press/v202/radford23a.html}.

\bibitem{qwen3tts}
\bibinfo{author}{{Qwen Team}}.
\newblock \bibinfo{title}{{Qwen3-TTS} ({Qwen3-TTS-12Hz-1.7B-VoiceDesign} and
  {Qwen3-TTS-12Hz-1.7B-Base})}.
\newblock
  \bibinfo{howpublished}{\url{https://huggingface.co/Qwen/Qwen3-TTS-12Hz-1.7B-VoiceDesign}}
  (\bibinfo{year}{2026}).
\newblock \bibinfo{note}{Model cards; accessed October 2026. Base model:
  \url{https://huggingface.co/Qwen/Qwen3-TTS-12Hz-1.7B-Base}}.

\bibitem{takamichi2020jvs}
\bibinfo{author}{Takamichi, S.} \emph{et~al.}
\newblock \bibinfo{title}{{JSUT} and {JVS}: Free {Japanese} voice corpora for
  accelerating speech synthesis research}.
\newblock \emph{\bibinfo{journal}{Acoust. Sci. Technol.}}
  \textbf{\bibinfo{volume}{41}}, \bibinfo{pages}{761--768}
  (\bibinfo{year}{2020}).
\newblock \urlprefix\url{https://doi.org/10.1250/ast.41.761}.

\bibitem{tomashenko2026voiceprivacy}
\bibinfo{author}{Tomashenko, N.} \emph{et~al.}
\newblock \bibinfo{title}{The third {VoicePrivacy} challenge: Preserving
  emotional expressiveness and linguistic content in voice anonymization}.
\newblock \emph{\bibinfo{journal}{Computer Speech \& Language}}
  \textbf{\bibinfo{volume}{100}}, \bibinfo{pages}{101988}
  (\bibinfo{year}{2026}).
\newblock \urlprefix\url{https://doi.org/10.1016/j.csl.2026.101988}.

\bibitem{bigdely2015}
\bibinfo{author}{Bigdely-Shamlo, N.}, \bibinfo{author}{Mullen, T.},
  \bibinfo{author}{Kothe, C.}, \bibinfo{author}{Su, K.-M.} \&
  \bibinfo{author}{Robbins, K.~A.}
\newblock \bibinfo{title}{The {PREP} pipeline: standardized preprocessing for
  large-scale {EEG} analysis}.
\newblock \emph{\bibinfo{journal}{Front. Neuroinform.}}
  \textbf{\bibinfo{volume}{9}}, \bibinfo{pages}{16} (\bibinfo{year}{2015}).
\newblock \urlprefix\url{https://doi.org/10.3389/fninf.2015.00016}.

\bibitem{robbins2020}
\bibinfo{author}{Robbins, K.~A.}, \bibinfo{author}{Touryan, J.},
  \bibinfo{author}{Mullen, T.}, \bibinfo{author}{Kothe, C.} \&
  \bibinfo{author}{Bigdely-Shamlo, N.}
\newblock \bibinfo{title}{How sensitive are {EEG} results to preprocessing
  methods: a benchmarking study}.
\newblock \emph{\bibinfo{journal}{IEEE Trans. Neural Syst. Rehabil. Eng.}}
  \textbf{\bibinfo{volume}{28}}, \bibinfo{pages}{1081--1090}
  (\bibinfo{year}{2020}).
\newblock \urlprefix\url{https://doi.org/10.1109/TNSRE.2020.2980223}.

\bibitem{pernet2019}
\bibinfo{author}{Pernet, C.~R.} \emph{et~al.}
\newblock \bibinfo{title}{{EEG-BIDS}, an extension to the brain imaging data
  structure for electroencephalography}.
\newblock \emph{\bibinfo{journal}{Sci. Data}} \textbf{\bibinfo{volume}{6}},
  \bibinfo{pages}{103} (\bibinfo{year}{2019}).
\newblock \urlprefix\url{https://doi.org/10.1038/s41597-019-0104-8}.

\bibitem{goncharova2003}
\bibinfo{author}{Goncharova, I.}, \bibinfo{author}{McFarland, D.},
  \bibinfo{author}{Vaughan, T.} \& \bibinfo{author}{Wolpaw, J.}
\newblock \bibinfo{title}{{EMG} contamination of {EEG}: spectral and
  topographical characteristics}.
\newblock \emph{\bibinfo{journal}{Clin. Neurophysiol.}}
  \textbf{\bibinfo{volume}{114}}, \bibinfo{pages}{1580--1593}
  (\bibinfo{year}{2003}).
\newblock \urlprefix\url{https://doi.org/10.1016/s1388-2457(03)00093-2}.

\bibitem{whitham2007}
\bibinfo{author}{Whitham, E.~M.} \emph{et~al.}
\newblock \bibinfo{title}{Scalp electrical recording during paralysis:
  quantitative evidence that {EEG} frequencies above 20~{Hz} are contaminated
  by {EMG}}.
\newblock \emph{\bibinfo{journal}{Clin. Neurophysiol.}}
  \textbf{\bibinfo{volume}{118}}, \bibinfo{pages}{1877--1888}
  (\bibinfo{year}{2007}).
\newblock \urlprefix\url{https://doi.org/10.1016/j.clinph.2007.04.027}.

\bibitem{maekawa2003}
\bibinfo{author}{Maekawa, K.}
\newblock \bibinfo{title}{Corpus of {Spontaneous} {Japanese}: its design and
  evaluation}.
\newblock In \emph{\bibinfo{booktitle}{Proceedings of the {ISCA}/{IEEE}
  Workshop on Spontaneous Speech Processing and Recognition ({SSPR} 2003)}}
  (\bibinfo{address}{Tokyo, Japan}, \bibinfo{year}{2003}).
\newblock
  \urlprefix\url{https://www.isca-archive.org/sspr_2003/maekawa03_sspr.html}.
\newblock \bibinfo{note}{Paper MMO2}.

\bibitem{naatanen1987}
\bibinfo{author}{N{\"a}{\"a}t{\"a}nen, R.} \& \bibinfo{author}{Picton, T.}
\newblock \bibinfo{title}{The {N1} wave of the human electric and magnetic
  response to sound: a review and an analysis of the component structure}.
\newblock \emph{\bibinfo{journal}{Psychophysiology}}
  \textbf{\bibinfo{volume}{24}}, \bibinfo{pages}{375--425}
  (\bibinfo{year}{1987}).
\newblock \urlprefix\url{https://doi.org/10.1111/j.1469-8986.1987.tb00311.x}.

\bibitem{crowley2004}
\bibinfo{author}{Crowley, K.~E.} \& \bibinfo{author}{Colrain, I.~M.}
\newblock \bibinfo{title}{A review of the evidence for {P2} being an
  independent component process: age, sleep and modality}.
\newblock \emph{\bibinfo{journal}{Clinical Neurophysiology}}
  \textbf{\bibinfo{volume}{115}}, \bibinfo{pages}{732--744}
  (\bibinfo{year}{2004}).

\bibitem{hickok2007}
\bibinfo{author}{Hickok, G.} \& \bibinfo{author}{Poeppel, D.}
\newblock \bibinfo{title}{The cortical organization of speech processing}.
\newblock \emph{\bibinfo{journal}{Nat. Rev. Neurosci.}}
  \textbf{\bibinfo{volume}{8}}, \bibinfo{pages}{393--402}
  (\bibinfo{year}{2007}).
\newblock \urlprefix\url{https://doi.org/10.1038/nrn2113}.

\bibitem{cheung2016}
\bibinfo{author}{Cheung, C.}, \bibinfo{author}{Hamilton, L.~S.},
  \bibinfo{author}{Johnson, K.} \& \bibinfo{author}{Chang, E.~F.}
\newblock \bibinfo{title}{The auditory representation of speech sounds in human
  motor cortex}.
\newblock \emph{\bibinfo{journal}{eLife}} \textbf{\bibinfo{volume}{5}},
  \bibinfo{pages}{e12577} (\bibinfo{year}{2016}).
\newblock \urlprefix\url{https://doi.org/10.7554/elife.12577}.

\bibitem{mcadam1971}
\bibinfo{author}{McAdam, D.~W.} \& \bibinfo{author}{Whitaker, H.~A.}
\newblock \bibinfo{title}{Language production: electroencephalographic
  localization in the normal human brain}.
\newblock \emph{\bibinfo{journal}{Science}} \textbf{\bibinfo{volume}{172}},
  \bibinfo{pages}{499--502} (\bibinfo{year}{1971}).
\newblock \urlprefix\url{https://doi.org/10.1126/science.172.3982.499}.

\end{thebibliography}

\end{document}